\documentclass[fleqn,10pt]{wlscirep}
\usepackage{amssymb,amsmath,float}
\usepackage{multirow}
\usepackage{bm}
\usepackage{epsfig}
\usepackage{multicol}        % used for the two-column index
\usepackage{url}
\usepackage{authblk}
\usepackage{bbold}
\usepackage{tabularx}
\usepackage{lscape}
\usepackage{graphicx}
\usepackage{enumitem}
\newcolumntype{b}{>{\hsize=1.45\hsize}X}
\newcolumntype{s}{>{\hsize=.15\hsize}X}
\newcolumntype{m}{>{\hsize=.7\hsize}X}
\usepackage{comment}

\begin{document}
\title{Dynamics of the market states in the space of correlation matrices with applications to financial markets}
%%%%%%%%%%%%%%%%%%%%%%%%%%%%%%%%%%%%%%%%%%%%%%%%%%%%%%%%%%%%%%%%%%%%%%%%%%%%%%%%%%%%
\author[1]{Hirdesh K. Pharasi}
\author[2]{Suchetana Sadhukhan}
\author[3]{Parisa Majari}
\author[4,5,6,7]{Anirban Chakraborti}
\author[3,7]{Thomas H. Seligman}
\affil[1]{Fakult\"at f\"ur Physik, Universit\"at Duisburg-Essen, Duisburg-47048, Germany}
\affil[2]{Department of Physics, School of Advanced Sciences, VIT Bhopal, Kothri Kalan, Sehore-466114, India}
\affil[3]{Instituto de Ciencias F\'{i}sicas, Universidad Nacional Aut\'{o}noma de M\'{e}xico, Cuernavaca-62210, M\'{e}xico}
\affil[4]{School of Engineering and Technology, BML Munjal University, Gurugram, Haryana-122413, India}
\affil[5] {School of Computational and Integrative Sciences, 
Jawaharlal Nehru University, New Delhi-110067, India}
\affil[6]{Centre for Complexity Economics, Applied Spirituality and Public Policy (CEASP), Jindal School of Government and Public Policy, O.P. Jindal Global University, Sonipat-131001, India}
\affil[7]{Centro Internacional de Ciencias, Cuernavaca-62210, M\'{e}xico}
\affil[*]{hirdeshpharasi@gmail.com; anirban.chakraborti@bmu.edu.in}
\date{\today}
%%%%%%%%%%%%%%%%%%%%%%%%%%%%%%%%%%%%%%%%%%%%%%%%%%%%%%%%%%%%%%%%%%%%%%%%%%%%%%%%%%%%
\begin{abstract}
%With a natural urge to understand the dynamics of complex systems, the analysis of correlations between constituents is crucial. Based on the similarity of the correlation matrices, a characterized “states” or “regions” is defined through which the system evolves over time. The concept of regions or states is not new but in the context of correlation similarities, the idea was first time introduced by Munnix et. al. in 2012. In this book chapter, we present briefly different approaches that are evolved to characterized market states  based on the similarity of the correlation matrices. We also present a cluster analysis of market states constructed from averaged intra- and inter-sectorial matrices to identify the market dynamics. The results are statistically similar to the one obtained from the stock analysis. It provides an important tool to understand the system in a low-dimensional space which reduces the complexity of the complex financial market. We then try to understand the system based on the occurrences of correlation structures as well as the dynamical evolution of trajectories in the space of states which reduces the complexity and provides a better understanding of the system.....

The concept of states of financial markets based on correlations has gained increasing attention during the last 10 years. We propose to retrace some important steps up to 2018, and then give a more detailed view of recent developments that attempt to make the use of this more practical. Finally, we try to give a glimpse to the future proposing the analysis of trajectories in correlation matrix space directly or in terms of symbolic dynamics as well as attempts to analyze the clusters that make up the states in a random matrix context.

\end{abstract}

%\pacs{89.65.Gh, 89.90.+n}

\maketitle
\section{Introduction}
The purpose of this chapter is to give an overview of recent developments in the evolution of the concept of `states' and `regions' of financial markets based on correlation analysis \cite{munnix2012identifying, rinn2015dynamics, pharasi2020dynamics}. Many refinements have been conceived, some of which try to account for a practical applicability of the concept \cite{munnix2012identifying,Pharasi_2018,Pharasi_2019,chetalova2015zooming,stepanov2015stability,
rinn2015dynamics, guhr2020exact,Heckens_2020,Meudt_2015,guhr2015non}. It opens a new perspective, namely the stochastic dynamics of markets in the space of correlation matrices; the dynamics are discrete in time by the nature of markets governed by individual and countable transactions. The space in which the correlation moves, is of high dimensionality and not a discrete grid! This continuous approach may also be relevant to other applications such as social systems, traffic, economy, ecological systems, neurology and many more\cite{husain2020identifying, wang2020quasi, scheffer_2009,scheffer_2012, May_2008, Sornette_2004,weiss2004social, wang2017unification,kawamura2012stat,muller2011evolution,kwapien2012physical}. 
Two points stand out in this context: On the one hand, the correlation matrices of financial markets have been studied over fairly long epochs, mostly with zero or 50\% overlapping epoch, which led to a low number of points for the time scale considered. On the other hand, the clusters which define market states are not obviously visible in the reduced spaces, where they have been visualized using multidimensional scaling (MDS) \cite{torgerson1952multidimensioal}. Smaller epoch shifts (bigger overlapping epochs) in time, to overcome the above mentioned reason as well as for practical purpose, indicate that a trader wants up to date analysis to make his decisions as pointed out in Ref. \cite{pharasi2020dynamics}. Using MDS, one can visualize the similar correlation matrices into 2D and 3D clusters but without a cleanly separated  boundaries. The transition matrices, consist of transition counts between the clusters, have shown their underlined relevance \cite{Pharasi_2018,Pharasi_2019,pharasi2020dynamics}. Therefore, following the trajectories on shorter time scales as well as introducing symbolic dynamics seems very attractive. The former is in an advanced stage and will be described in this chapter in some detail, while the latter will be proposed and potential usefulness will be pointed out. In both cases, practically useful results are still outstanding. In the framework of these ideas, we start with a brief description of the origins of the correlation analysis and its evolution. We show how this naturally leads to an evolution in a space, whose dimension is determined by the length of the time series considered, i.e., by the time horizon as well as the number of free matrix elements in the correlation matrix. The analysis, based on the distances in the correlation matrix space between the points touched in time, defines the trajectory and the geometrical structure of the trajectory. Studying the correlation of market sectors following a technique proposed in Ref. \cite{rinn2015dynamics} yields similar results as we shall see. The high dimensionality of the space in practice makes visualization very difficult.

We could try to represent each state by, say, its average correlation matrix and this would reduce the problem to a discrete space as we would identify all correlation matrices in a cluster with a single label or symbol. At this point it should be noticed that for traffic a three phase theory\cite{kerner2009introduction} is used; similarly, a four market states model is possible for the S\&P 500 market and a higher number of states is certainly required for the Nikkei 225 market due to higher complexity of the Japanese market\cite{Pharasi_2018}. Furthermore in the Japanese market a dominance of the average correlation for the structure of market states, found in the S\&P 500 market \cite{munnix2012identifying}, clearly does not hold. This opens a perspective for the search of a second relevant parameter, which we shall not pursue here.

In the framework of published and submitted works, which we will briefly describe, we propose two routes to advance the use of correlation matrices and market states. The first is MDS, say, to two or three dimensions, and the second is a symbolic dynamics after a partitioning of the set of observed correlation matrices according to the cluster structure used for defining the states.  It, in turn, partitions the space on which the symbolic dynamics evolve in time.
We shall show that in the two relevant financial markets (S\&P 500 and Nikkei 225) {this structure is only marginally affected by MDS}, which thus emerges as  a very effective tool for visualization and interpretation. We will barely outline the second option, namely the possible usefulness of the symbolic dynamics resulting from the developed cluster structure. In financial markets, we have previously opted epochs shifted by one day \cite{pharasi2020dynamics}, thus omitting the intra day trading data, both because they are more difficult to access and of different nature than daily trading data. We limit our study to a time interval 2006-2019 excluding the year 2020, though we will show scaled trajectory for 2020 that is interesting yet lack of interpretation at present.

We shall develop basics and outline the development that led over the years to the present work. Next we give some technical aspects of data handling and processing. Then we proceed to show results for the correlations of the USA market as reflected by the S\&P 500 and the Tokyo Market by the Nikkei 225 using both full correlation matrices and market sector averages. We start by discussing the market states, their transition matrices and then passing to the trajectories in the space of correlation matrices with visualizations including at this point some aspects of the 2020 development of markets during the starting period of COVID-19 pandemic. At this point we shall also mention trajectories in the symbolic dynamics. No conclusions in either case can be presented as it is an ongoing work. Finally we shall try to give an outlook as to where we think the {field} is heading and what could be done about applications to other fields.

\section{Methodology}
\subsection{Data description}\label{Sec:Materials}
We analyze the daily adjusted closure prices $S_k(t)$ of stock $k$ for the S\&P 500 and Nikkei 225 indices at different time horizons. The daily trading data is freely available at Yahoo finance website\cite{Yahoo_finance}. We consider only those stocks  which are continuously traded over the considered time period. We again filtered out a few stocks time series that have more than two consecutive missing trading days entries. The price value for the missing days has chosen to be the same as the one on the previous day entry. We then calculate the logarithmic price returns $r_k(t) =ln \; S_k(t+ \Delta t)-ln \; S_k(t)$  for each stock $k$ at time $t$, where $t = 1, 2,\dots,T_{tot}$ with $T_{tot}$ is total number of the trading days present in the time horizon considered, and $k=1,\dots,N$ for $N$ number of stocks and $\Delta t=1$ day for daily price returns. 

\subsection{Evolution of cross-correlation structures}
To understand the co-movement of the financial market constituents (stocks) and the time evolution of the market conditions, we calculate the Pearson correlation coefficients $C_{ij}(\tau)=({\langle r_ir_j\rangle -\langle r_i \rangle\langle r_j \rangle})/{\sigma_{r_i}\sigma_{r_j}}$ where $i,j=1,\dots,N$, $\tau$ is the epoch number of size $T$ trading days and $\sigma$ is the standard deviation. Here $\langle ... \rangle$ denotes an averaging over an epoch length of $T$ trading days which is shifted by $\Delta T$ days through the data. 
The choice of epoch is important because market conditions evolute with time and the cross-correlations between any pair of stocks may not be stationary for a long epoch, on the other hand, short epoch size leads to ``noise'' or ``fluctuation'' due to the singularity present in the corresponding correlation matrix.  Therefore, the distribution of eigenvalue spectrum of the empirical cross-correlation matrix $\boldsymbol C (\tau)$ contains ``random'' and as well as non-random contribution \cite{plerou2000random, pandey2010correlated}. That motivates us to compare the eigenvalue statistics of $\boldsymbol C (\tau)$ with a large random correlation ensemble constructed from mutually uncorrelated time series or white noise known as Wishart matrix. We will introduce Wishart Orthogonal Ensembles (WOE) in detail in the next subsection. 

\subsection{Wishart Orthogonal Ensembles}
We construct a rectangular random data matrix ${\boldsymbol A}= [A_{ij}]$ of $N$ random time series each of length $T$, i.e., of order $N \times T$ with real independent elements drawn from a standard Gaussian distribution with fixed mean and  variance. The Wishart matrix is then constructed as
${\boldsymbol W}=\frac{1}{T}{\boldsymbol  A}{\boldsymbol A}'$ of size $N \times N$, where ${\boldsymbol A}'$ denotes the transpose of the matrix ${\boldsymbol A}$. Wishart matrices are random matrix models used to describe universal features of covariance matrices. We here consider the entries as real, known in the literature as Wishart orthogonal ensemble and also, by construction, it is a real symmetric positive semidefinite matrix. To obtain the correlation matrix, we fix the value of mean to zero and variance to $\sigma^2=1$. In the context of time series, ${\boldsymbol W}$ may be interpreted as the correlation matrix, calculated over stochastic time series of time horizon $T$ for $N$ statistically independent variables. Following the Mar\u{c}enko-Pastur distribution, the probability density function $\bar{\rho}(\lambda)$ of the eigenvalues of such correlation matrices can be analytically shown for the limit $N \rightarrow \infty $ and $T \rightarrow \infty$ as 

$\bar{\rho}(\lambda)=\left\{ \begin{array}{rcl} \frac{Q}{2\pi \sigma^2}{\sqrt{(\lambda_{max}-\lambda)(\lambda-\lambda_{min})}}/{\lambda} &\text{if }& Q> 1\\
\frac{Q}{2\pi \sigma^2}{\sqrt{(\lambda_{max}-\lambda)(\lambda-\lambda_{min})}}/{\lambda} + (1-Q)\delta(\lambda) &\text{if }& Q \leq 1
\end{array}\right.$\\
where $Q = T /N$  fixed. The maximum and minimum eigenvalue can be derived 
from $ \lambda_{min}^{max}= \sigma^2 \left(1\pm  {1}/{\sqrt{Q}}\right)^2$. For  $Q \leq 1$, one has to take into account of at least $N-T$ zeros, hence the density $\bar{\rho}(\lambda)$ is normalized to $Q$ and not to unity. 

\subsection{Power map technique}
To obtain the stationarity in time series, the consideration of shorter time series by breaking a long time series into epochs, each of size $T$ (such that $T_{tot}/T=Fr$) are useful. However, if $N$ (stocks)$ > T$, we get a singular correlation matrix with $N-T+1$ zero eigenvalues, results in poor eigenvalue statistics. We introduce here the power map technique \cite{Guhr_2003, vinayak_2014} to lift the degeneracy of zero eigenvalues by providing a non-linear distortion ($\epsilon$, also called noise suppression parameter) to each element $(W_{ij})$ of the Wishart matrix $\boldsymbol W$: $
W_{ij} \rightarrow (\mathrm{sign} ~W_{ij}) |W_{ij}|^{1+\epsilon} $. This technique helps to get an ``emerging spectrum'' of eigenvalues by removing the degenerated eigenvalues at zero even for very small distortions, e.g., $\epsilon=0.001$ (see e.g., Refs.~\cite{Chakraborti_2020, Pharasi_2018, vinayakpre2013} for recent studies and applications). 
%%%%==============  Figure 1  =========%%%%%%%%%%%%%%%%%%%%%
\begin{figure}[b!]
\centering
\includegraphics[width=1\linewidth]{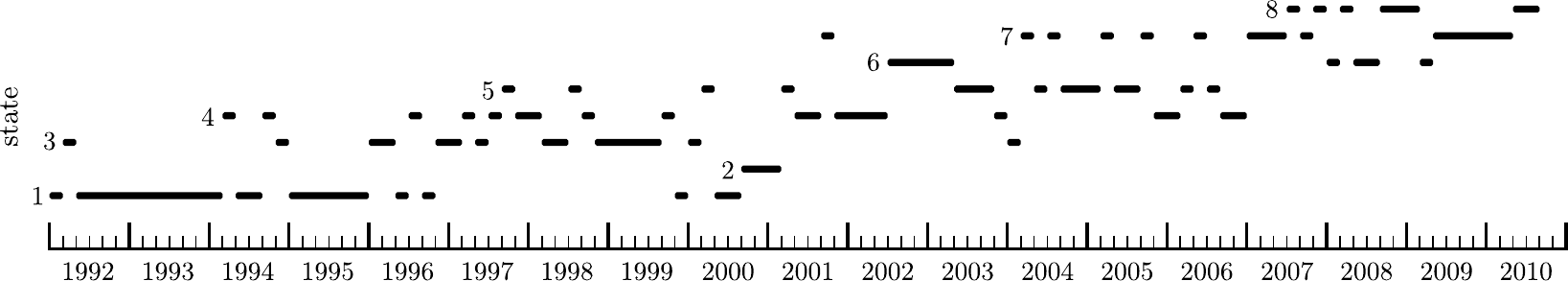}
\caption{The plot shows the dynamical evolution of the market states for S\&P 500 market over the 19-year period 1992-2010. The market shows the transitions between eight characterized states (S1 to S8) based on the top-down clustering method. The lower states (i.e., S1 and S2), denoting as normal periods of the market, tend to cluster in time revealing that the market remains in those particular states for a longer time. On the other hand, higher states are scattered and irregular in time showing the rapid to and fro transitions between the states with the higher probability of transitions to the nearby states. Figure is adapted from the Ref \cite{munnix2012identifying}.}\label{fig_2012}
\end{figure}
%%%%%%%%%%%%%%%%%%%%%%%%%%%%%%%%%%%%%%%%%%%%%%%%%
\subsection{Pairwise (dis)similarity measures and Multidimensional scaling}
In pairwise (dis)similarity analysis, the correlation matrix is first calculated for each epoch (which represents a specific time frame) and we compute the (dis)similarity by measuring the distance between the two correlation matrices. Such analysis is performed for each pair of correlation matrices (frames) and defined as:
$\zeta(\tau_1, \tau_2) \equiv \langle | C_{ij} (\tau_1)-  C_{ij} (\tau_2)|  \rangle$, where $|...|$ and $\langle ... \rangle$ denote the absolute and average value over all the matrix elements, respectively and $\tau=1,\dots,Fr$ with $Fr$ is the total number of epochs of size $T$ constructed from the overlapping shift of $\Delta T$ days. This technique yields a symmetric matrix, with all positive off-diagonal elements $Fr(Fr-1)/2$, where $Fr$ is the number of frames. Diagonal elements are zero for no such (dis-)similarities present between the same frames. We use multidimensional scaling \cite{torgerson1952multidimensioal},  which reveals the structure of (dis-)similarity data by creating a map, it constructs a geometrical representation from the relative distances ($\zeta$) of correlation frames in such a way that the points corresponding to similar objects are located close together, while those corresponding to dissimilar objects are located far apart.

\section{Identifying states of a financial market}
In this section, we present an overview of recent developments in the evolution of the concept of `states' of financial stock markets based on correlation analysis.  We start with the very first work on the characterization of ``market states" based on the similarity of the correlation of market states by Munnix et. al. \cite{munnix2012identifying}. Figure \ref{fig_2012} shows the temporal evolution of the market states of the S\&P 500 market for 19-year observation period 1992–2010. Each point in the figure represents a correlation matrix measured over the previous two months. The market is clustered into eight states (S1 to S8) based on a top-down clustering method using a similarity matrix $\zeta$. {In this method, initially all the correlation matrices are considered as a single cluster and then further divided into sub-clusters using $k$-means clustering algorithm. The division process stops as the average radii of all the points to centroids becomes smaller than a certain threshold.} In the Fig. \ref{fig_2012}, initial points are clustered in lower states (S1 and S2) with less transitions to other states, hence shows a stable behavior of the market before 1996. On the other hand, one can observe more frequent jumps for higher states with scattered points after 1996 and shows high market volatility.  In this method, on one side, the number of clusters is dependent on an arbitrary threshold and on the other side, the correlation matrices constructed by short time series are highly singular and dressed with noise.

%%%%%%%%%%%%%%%%%%%%%%%%%%%%%%%%%%%%%%%%%%%%%%%%%%%%%%%%%%%%
%%%%%%%%%%%%%%%%%%%%%%%%%%%%%%%%%%%%%%%%%%%%%%%%%%%%%%%%%%%%
\begin{figure}[t!]
\centering
\includegraphics[width=.495\linewidth]{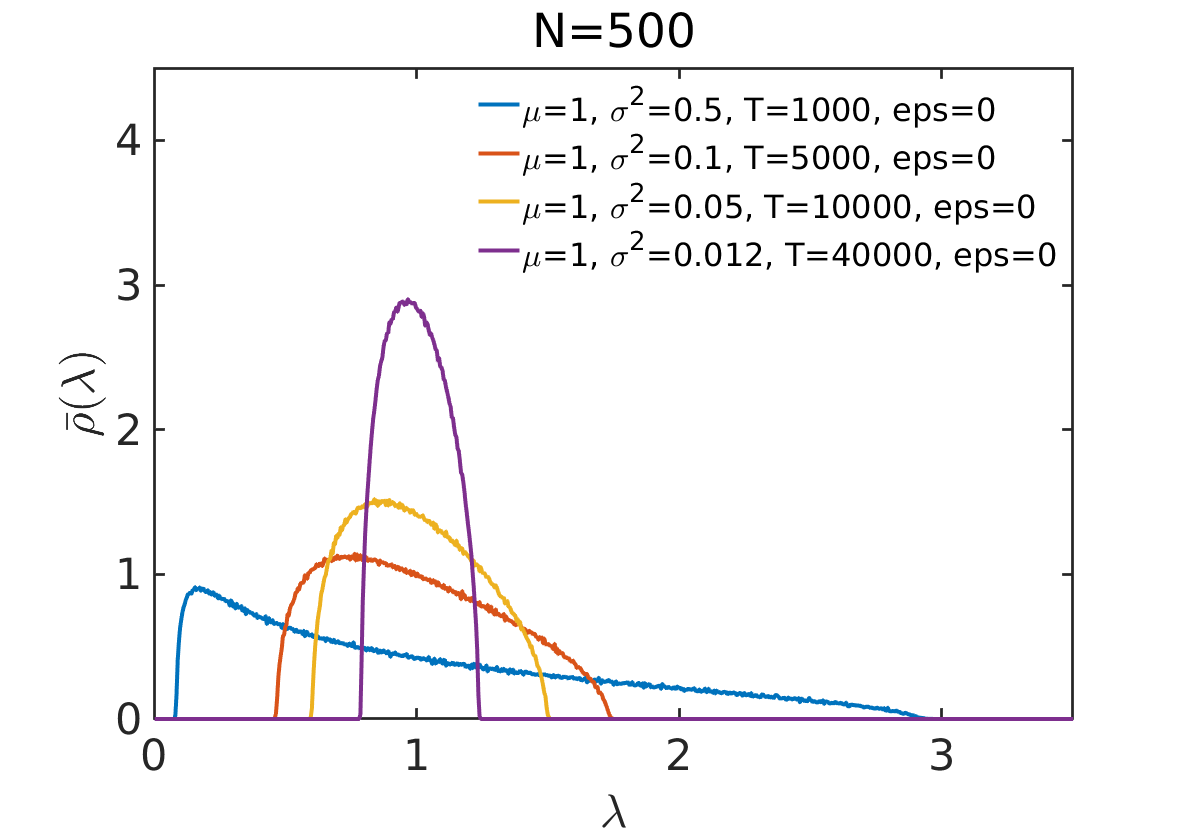}\llap{\parbox[b]{3.4in}{\textbf{\Large (a)}\\\rule{0ex}{2.4in}}}
\includegraphics[width=.495\linewidth]{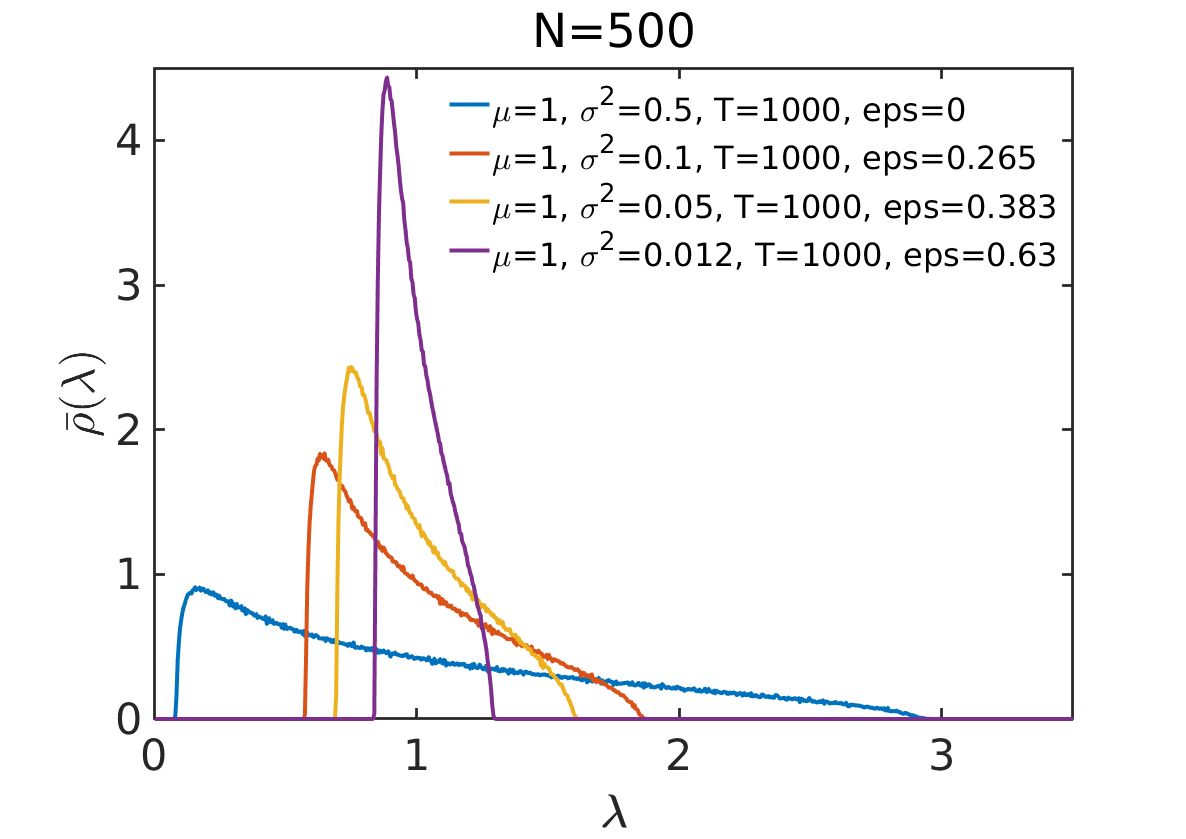}\llap{\parbox[b]{3.4in}{\textbf{\Large (b)}\\\rule{0ex}{2.4in}}}
\caption{Effect of the power map method and length of time series on the spectral distribution of the Wishart orthogonal ensemble (WOE). (a) shows the effect of finite size on the spectral densities $\overline{\rho} (\lambda)$ of a Wishart matrix ensemble (1000) of size $N\times N$, constructed from N time series ($N=500$, fixed) of real independent Gaussian variables, each of finite epoch length $T$ (varying) with unit mean and varying variance $\sigma^2$. The variance of the distribution decreases with the increment of T and the ensemble is becoming an identity for higher $T$. The same can be achieved by increasing the value of the noise suppression parameter $\epsilon$ of the power map. (b) shows the effect of the power map by varying $\epsilon$ and keeping $N$ and $T$ fixed.}\label{fig_woe}
\end{figure}
%%%%%%%%%%%%%%%%%%%%%%%

The noise dressing has an enormous effect on portfolio optimization and proper risk assessment \cite{schafer2010power}. It has been found by Laloux et. al. \cite{laloux1999noise} that the correlation matrices are noisy due to the finiteness of the time series. Guhr et. al. \cite{guhr2003} has presented the power map method to suppress the noise of the correlation matrices, which effectively increases the length of time series. This method is more suitable for the analysis of short time series. In Fig. \ref{fig_woe}, we show the effect of the power map method and the length of time series on the spectral density $\overline{\rho} (\lambda)$ of the Wishart orthogonal ensemble (WOE). We use various epoch lengths $T=1000, 5000, 10000, 40000$ and noise suppression parameters $\epsilon=0, 0.265, 0.383, 0.63$ and show that the spectral density $\overline{\rho} (\lambda)$ of WOE approaches towards identity matrix for higher value of $T$ (for fixed $N$ and $\epsilon$) and $\epsilon$ (for fixed $N$ and $T$). Fig. \ref{fig_woe} (a) shows the spectral density $\overline{\rho} (\lambda$) of a Wishart matrix ensemble (ensemble $=$ 1000) of size $N\times N$, constructed from  $N=500$ (fixed) time series of real independent Gaussian variables, for four different epoch lengths $T$($1000, 5000, 10000, 40000$) with unit mean and variance $\sigma^2$. The variance of the distribution decreases with the increment of $T$ and move towards the identity matrix for higher $T$. The same can be achieved by increasing the value of the noise suppression parameter $\epsilon$ of the power map. Fig. \ref{fig_woe} (b) shows the effect of the power map by varying the value of $\epsilon$ but keeping $N$ and $T$ fixed. The comparison of spectral densities $\overline{\rho} (\lambda)$ in Fig. \ref{fig_woe} (a) and (b)  show that the similar variance can be achieved either by increasing the length of the time series or by increasing the value of noise suppression parameter $\epsilon$.

%%%%%%%%%%%%%%%%%%%%%%%%%%%%%%%%%%%%%%%%%%%%%%%%%%%%%%%%%%%%%%%
%%%%%%%%%%%%%%%%%%%%%%%%%%%%%%%%%%%%%%%%%%%%%%%%%%%%%%%%%%%%%%%%
\begin{figure}[ht!]
\centering
\includegraphics[width=.445\linewidth]{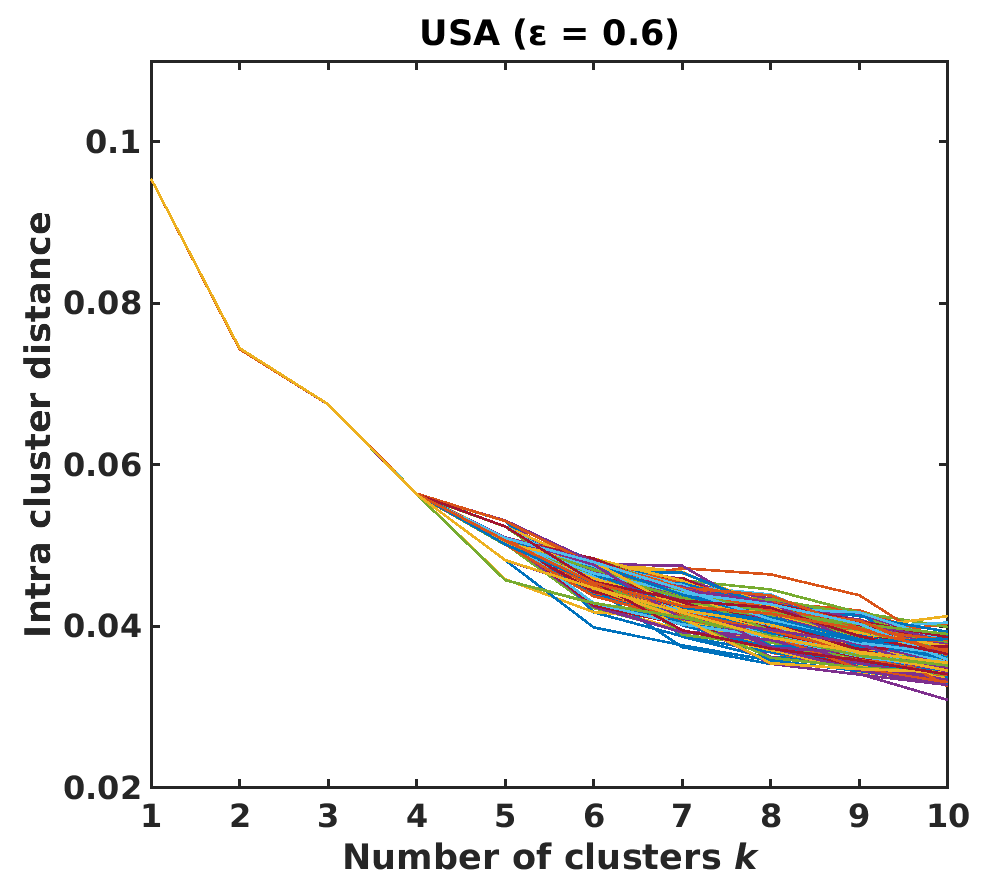}\llap{\parbox[b]{3.2in}{\textbf{\Large (a)}\\\rule{0ex}{2.4in}}}
\includegraphics[width=.445\linewidth]{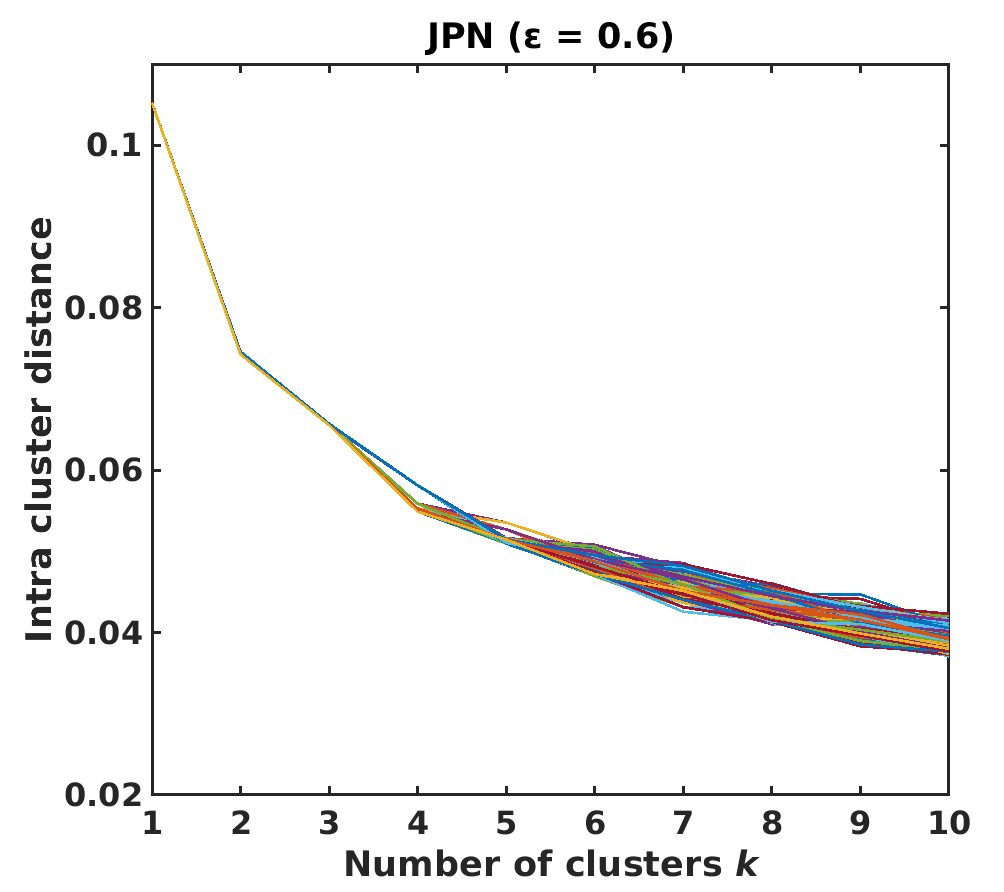}\llap{\parbox[b]{3.2in}{\textbf{\Large (b)}\\\rule{0ex}{2.4in}}}
\includegraphics[width=.445\linewidth]{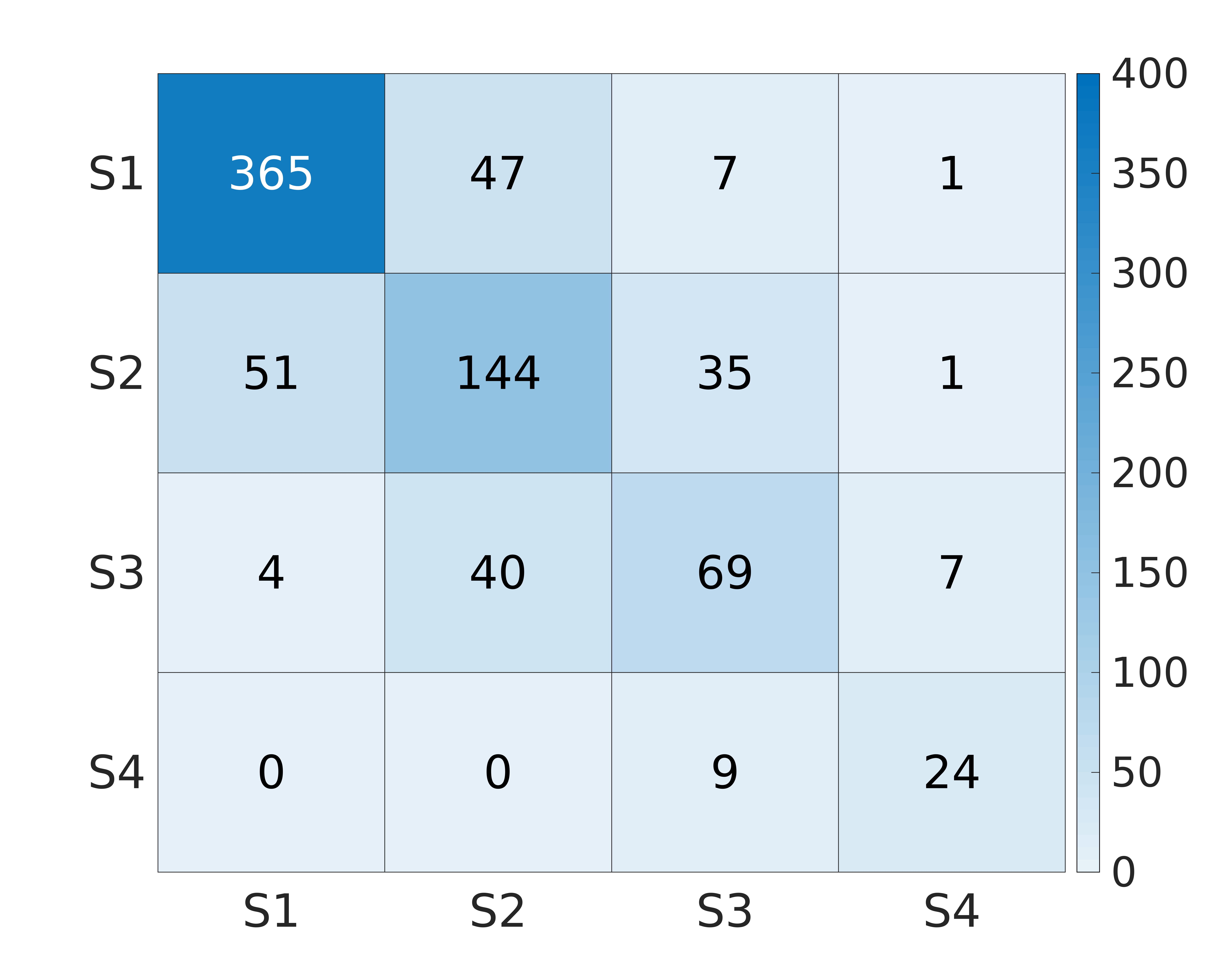}\llap{\parbox[b]{3.2in}{\textbf{\Large (c)}\\\rule{0ex}{2.4in}}}
\includegraphics[width=.445\linewidth]{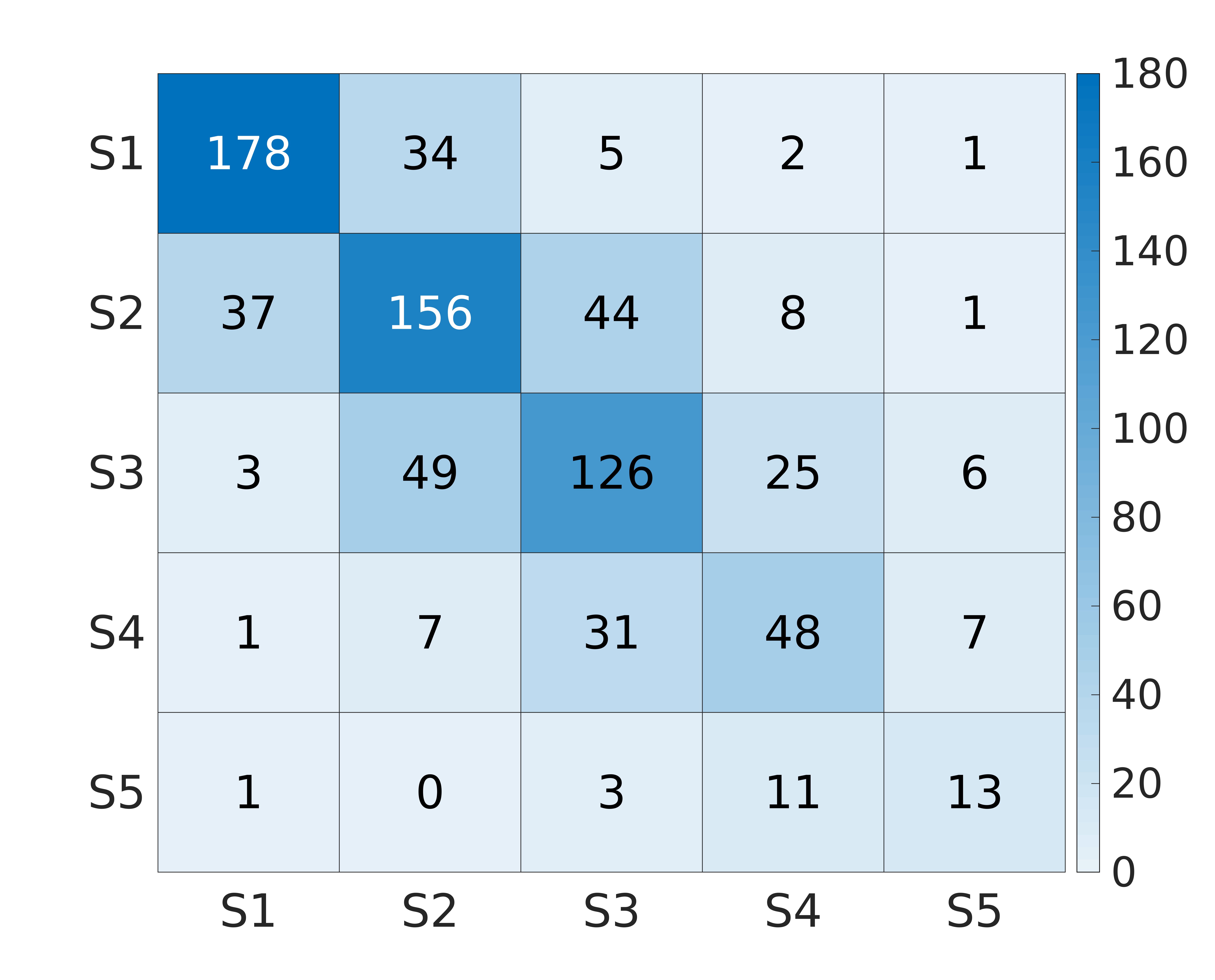}\llap{\parbox[b]{3.1in}{\textbf{\Large (d)}\\\rule{0ex}{2.4in}}}
\caption{Plots of (top row) intracluster distance as a function of the number of clusters and (bottom row) transition counts of paired market states for S\&P 500 ((a) and (c)) and Nikkei 225 ((b) and (d)) markets of noised suppressed ($\epsilon=0.6$)  correlation matrices constructed from the short epoch of length $T= 20$ days with shifts of $\Delta= 10$ days over the period 1985-2016. Five hundred different initial conditions are used to calculate the intracluster distances denoted by colored lines in (a) and (b). Based on the optimization conditions of keeping the standard deviation of the ensemble lowest and the number of clusters highest, we have found $k=4 $ and $k=5$ are the optimum number of clusters for the USA and JPN, respectively. Color map plots for (c) S\&P 500 and (d) Nikkei 225 markets show the transition counts of paired market states. In both markets, the highest transition counts in the diagonal show high re-occurrence of the market states followed by first off diagonal transitions to nearest states. Thus the penultimate state to the critical state behaves like a precursor to the market crash and is the best state to hedge against the crash. The figures are recreated based on the Ref.\cite{Pharasi_2018}} \label{fig_2018njp}
\end{figure}
%%%%%%%%%%%%%%%%%%%%%%%%%%%%%%%%%%%%%%%%%%%%%%%%%%%%%%%%%%%%%%%%%%%%%%
%%%%%%%%%%%%%%%%%%%%%%%%%%%%%%%%%%%%%%%%%%%%%%%%%%%%%%%%%%%%%%%%%%%%%%%%
\begin{figure}[htb!]
\centering
\includegraphics[width=.48\linewidth]{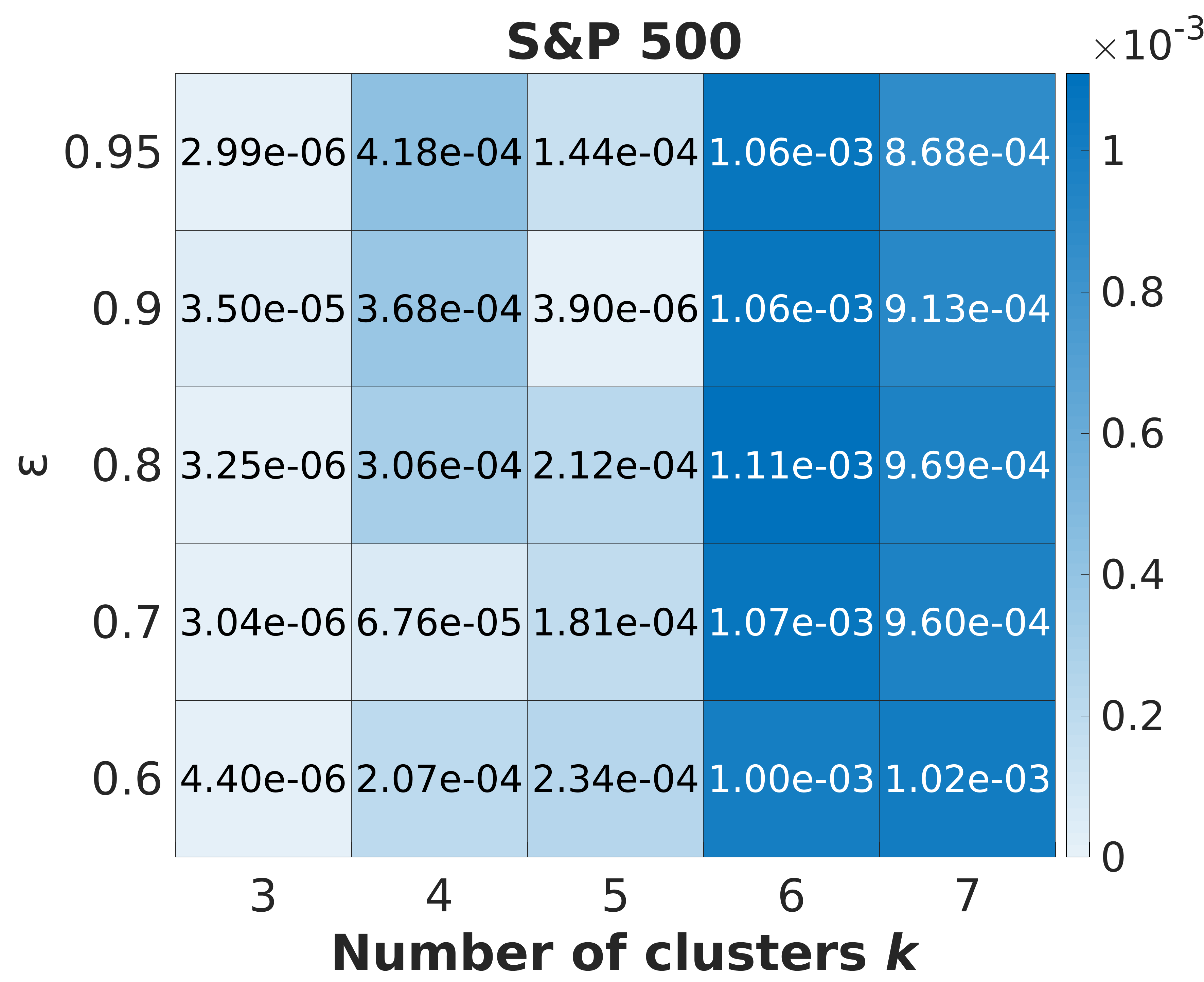}\llap{\parbox[b]{3.4in}{\textbf{\Large (a)}\\\rule{0ex}{2.5in}}}
\includegraphics[width=.48\linewidth]{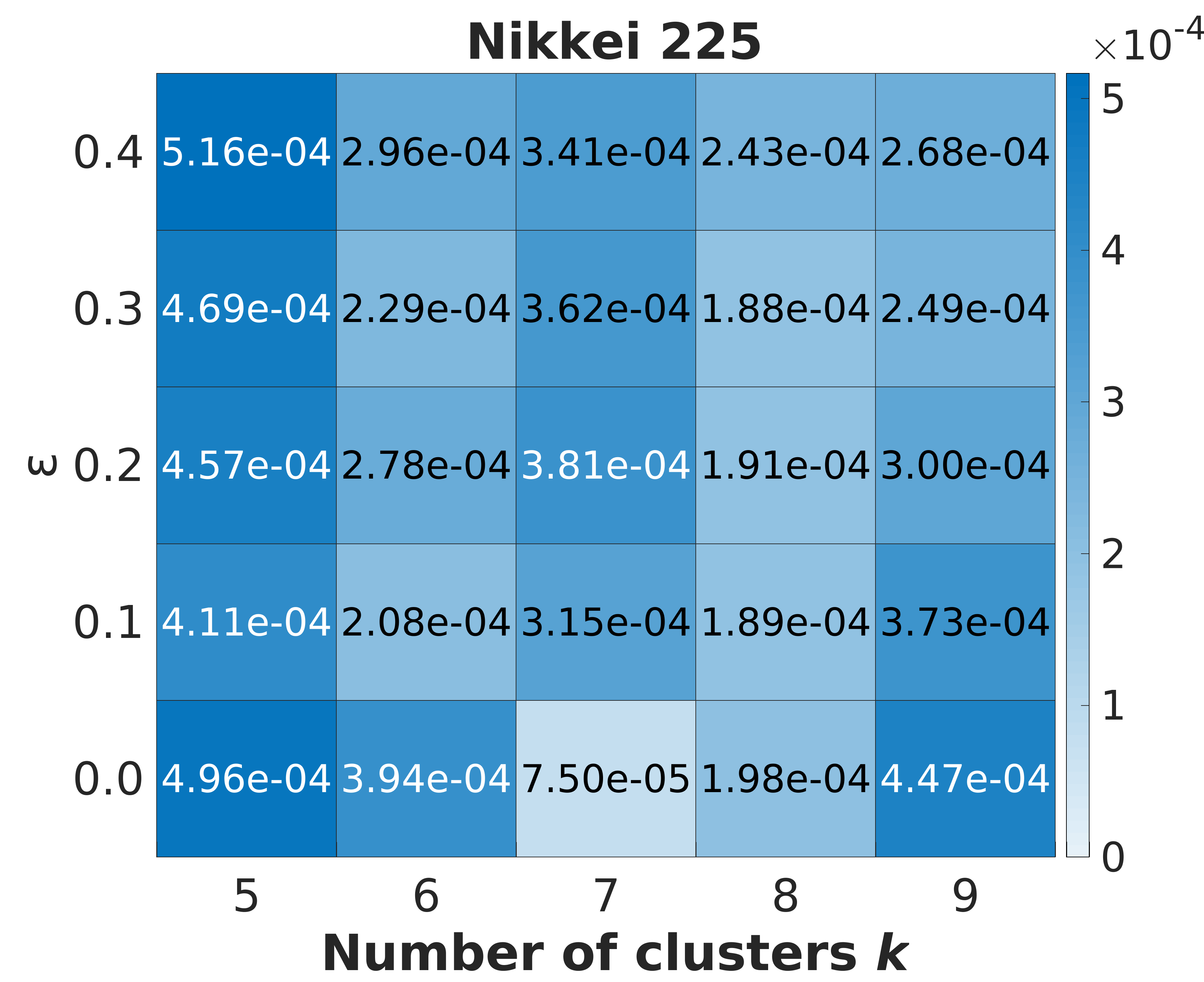}\llap{\parbox[b]{3.3in}{\textbf{\Large (b)}\\\rule{0ex}{2.5in}}}
\includegraphics[width=.48\linewidth]{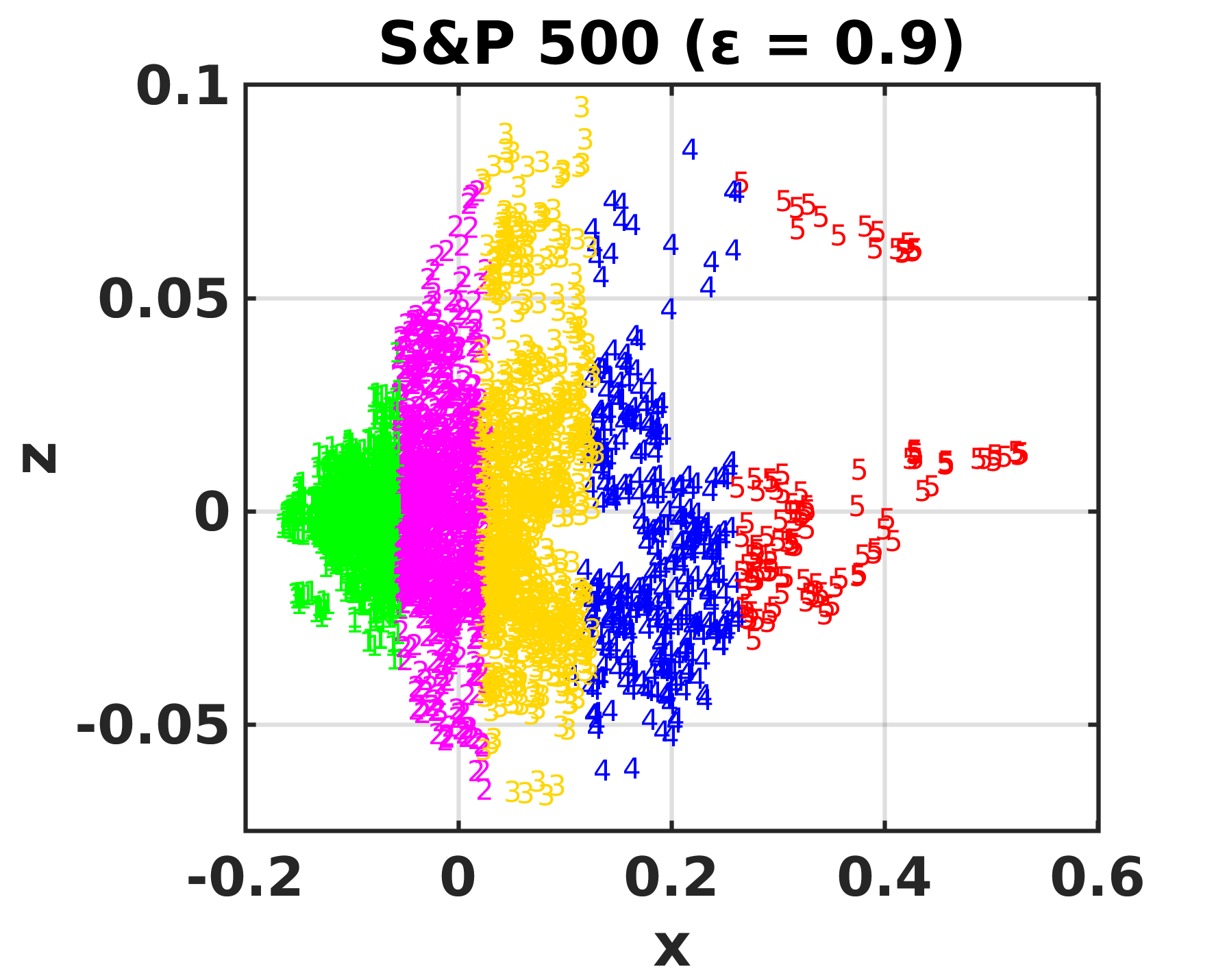}\llap{\parbox[b]{3.4in}{\textbf{\Large (c)}\\\rule{0ex}{2.7in}}}
\includegraphics[width=.48\linewidth]{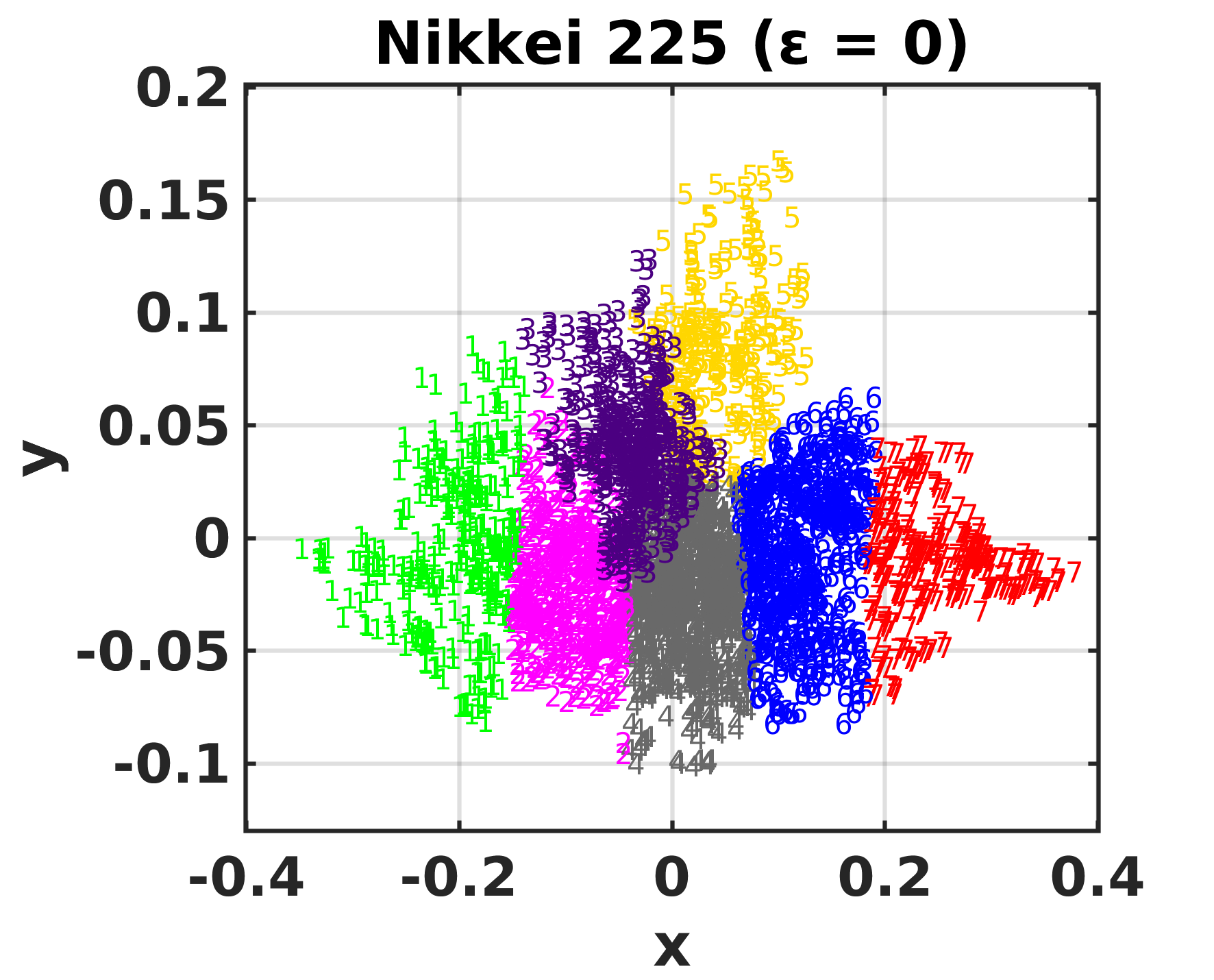}\llap{\parbox[b]{3.3in}{\textbf{\Large (d)}\\\rule{0ex}{2.7in}}}
\caption{Identification of number of market states based on minimum standard deviation of intra cluster distances $\sigma_{d_{intra}}$ (colorbar) for (a) S\&P 500 and (b) Nikkei 225 markets. For the correlation analysis considered in Ref. ~\cite{pharasi2020dynamics}, $N=350$ stocks of the S\&P 500 index and $N=156$ stocks of the Nikkei 225 index traded over 14-year period from 2006-2019 for overlapping epochs of length $T=20$ days shifted of $\Delta=1$ day are used. Plots show the measure of $\sigma_{d_{intra}}$ (colorbar) as a function of noise suppression parameter $\epsilon$ (zoomed in portion of Fig. 1 in Ref\cite{pharasi2020dynamics}). We have found that the minimum of $\sigma_{d_{intra}}$ calculated over 1000 initial conditions for $k\geq 4$ at $k=5$ and $\epsilon=0.9$ for S\&P 500 and at $k=7$ and $\epsilon=0$ for Nikkei 225 markets. Plots (c) and (d) show the $2D$ projection of $3D$ $k$-means clustering for S\&P market on $xz$-plane and Nikkei 225 market on $xy$-plane, respectively. It is important to note that the cluster distributions (only horizontal in the USA but both horizontal and vertical in JPN) for the two markets are significantly different. The figure is adapted from Ref. \cite{pharasi2020dynamics}.}\label{fig_MS2020_epsVsk}
\end{figure}
%%%%%%%%%%%%%%%%%%%%%%%%%%%%%%%%%%%%%%%%%%%%%%%%%%%%%%%%%%%%%%%%%%%%%%%%%%%%%%%%%%
%%%%==================  Figure 2  ====================%%%%%
\begin{figure}[ht!]
\centering
\includegraphics[width=.93\linewidth]{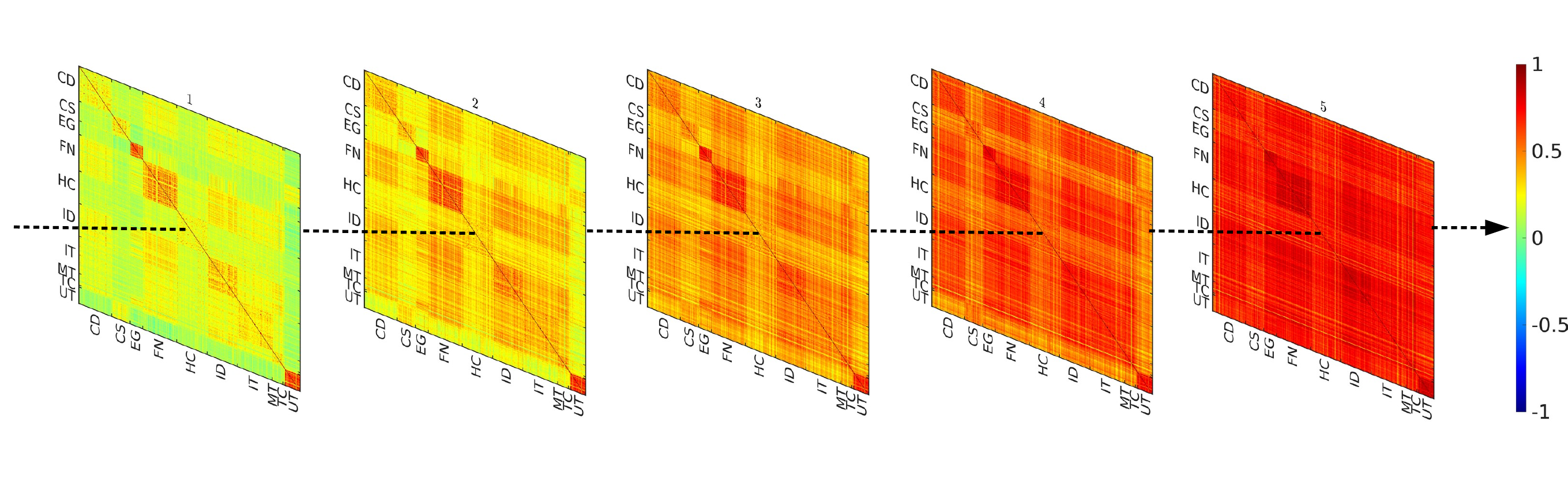}\llap{\parbox[b]{6.5in}{\textbf{\Large (a)}\\\rule{0ex}{1.6in}}}
\includegraphics[width=.93\linewidth]{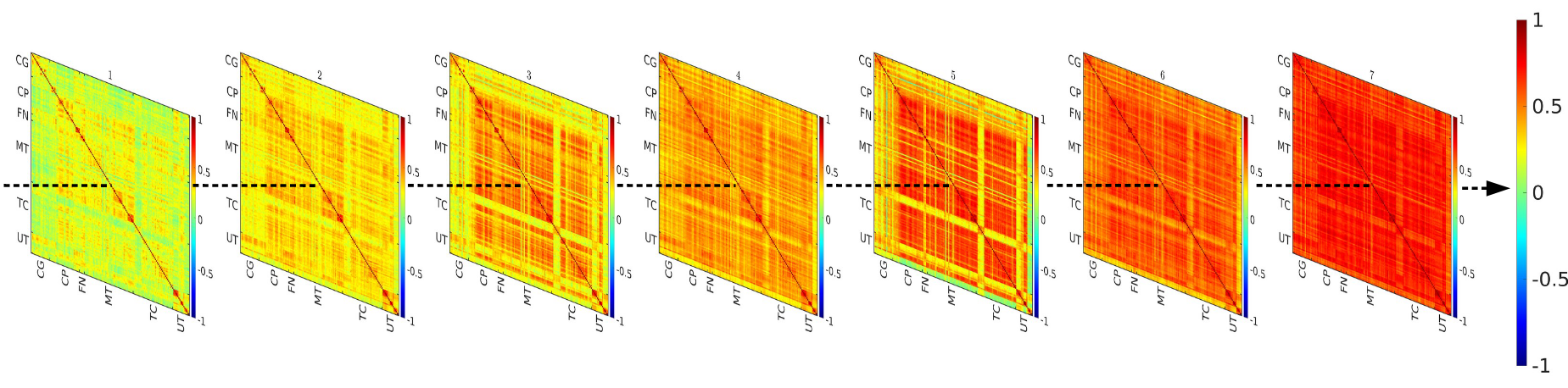}\llap{\parbox[b]{6.5in}{\textbf{\Large (b)}\\\rule{0ex}{1.4in}}}
\includegraphics[width=.48\linewidth]{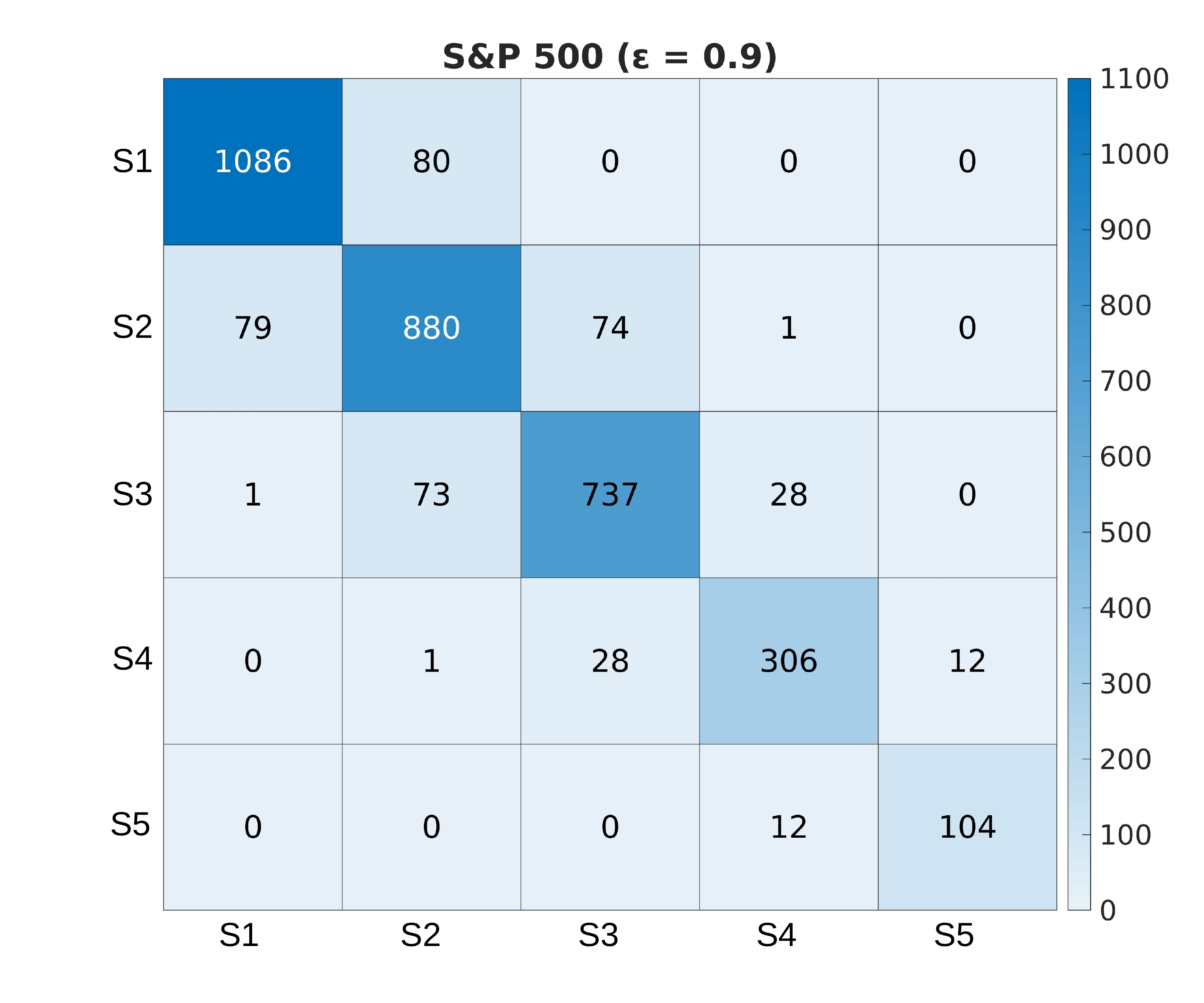}\llap{\parbox[b]{3.2in}{\textbf{\Large (c)}\\\rule{0ex}{2.5in}}}
\includegraphics[width=.48\linewidth]{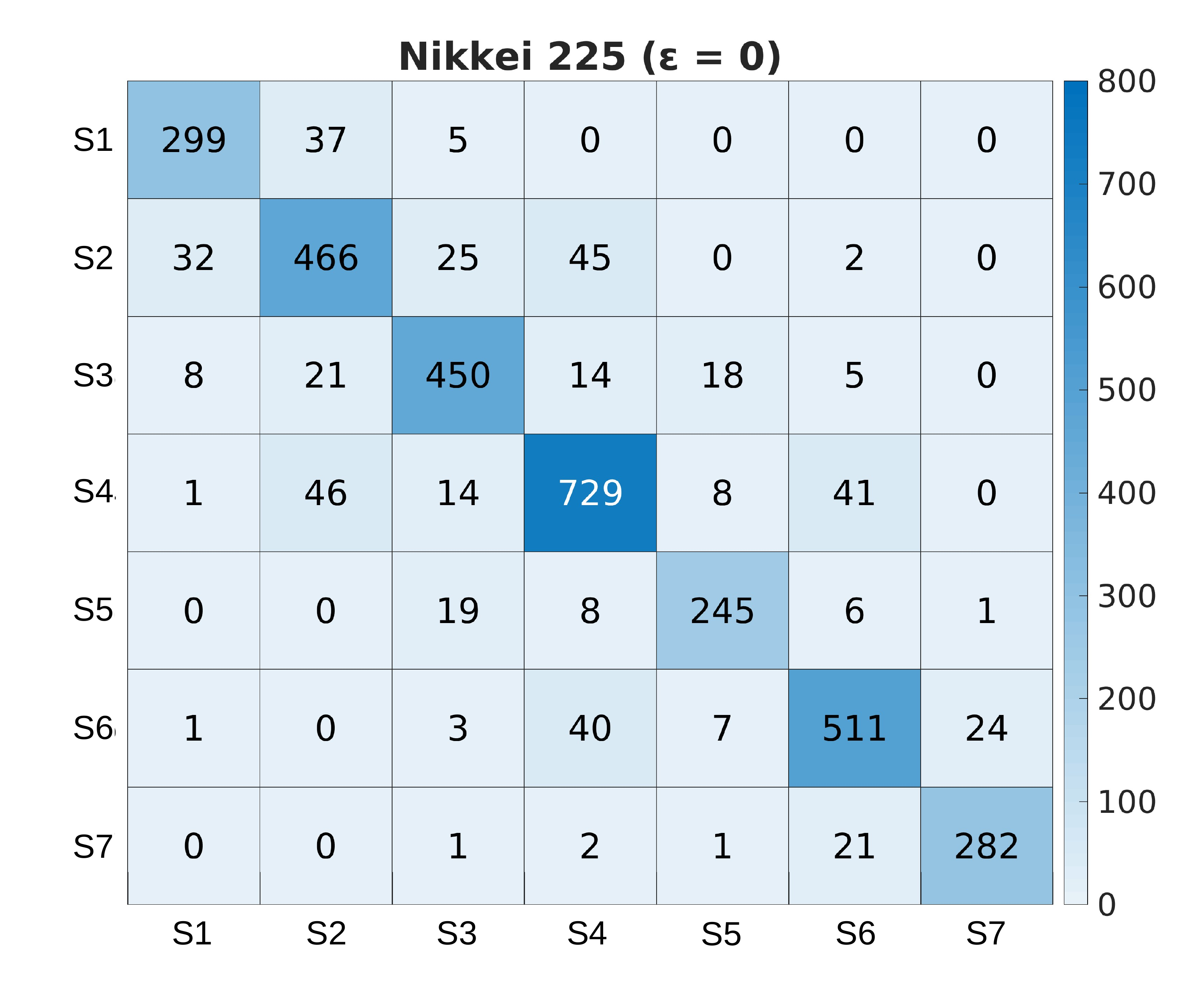}\llap{\parbox[b]{3.3in}{\textbf{\Large (d)}\\\rule{0ex}{2.5in}}}
\caption{Schematic diagram of average correlation matrices of (a) five market states of S\&P 500 and (b) and seven market states of Nikkei 225 markets  over the period of 2006-2019. The average is taken over all the correlation matrices of each market state. The correlation structure as well as mean correlations varies over different market states. Color map plots show the transition counts of paired market states (MS) for (c) S\&P 500 with $k=5$ clusters and noise-suppression parameter $\epsilon=0.9$ and (d) Nikkei 225 market with $k=7$ clusters and without noise-suppression ($\epsilon=0$). The transition matrix for (b) S\&P 500 market  is nearly tridiagonal and shows zero transitions from lower states (S1, S2, and S3) to critical state (S5) with twelve transitions from the penultimate state (S4) to critical state (S5). Thus, S4 behaves as a precursor to the market crashes and a good agent to hedge. For (d) Nikkei 225 market, the transition counts of the paired market states between S2 and S3 are smaller than S2 and S4. Here, state S6 behaves as a precursor to the critical state S7. The figure is adapted from Ref. \cite{pharasi2020dynamics}.}\label{fig_avg_corr_mat_MS}
\end{figure}

To suppress the noise of the correlation matrices and to avoid the arbitrariness of the threshold in the top-bottom clustering technique \cite{munnix2012identifying}, Pharasi et. al. \cite{Pharasi_2018} developed a new approach to obtain the optimum  number of market states for a financial market. Similar approach can be applied to other complex systems. The optimization is achieved using two parameters- intra cluster measure and noise suppression parameter. Intra cluster measure ${d_{intra}}$ is an averaged euclidean distance of points to cluster centroids of the $k$-means clustering performed on the MDS map. It is constructed from the similarity matrix $\zeta$ using 805 noise-suppressed ($\epsilon=0.6$) correlation matrices of USA market and 798 noise-suppressed ($\epsilon=0.6$) correlation matrices of Japanese market  and overlapping epoch $T= 20$ days shifted by $\Delta= 10$ days over the period 1985-2016, shown in Figs. \ref{fig_2018njp} (a) and (b), respectively. Five hundred different initial conditions are used to calculate the intra cluster distances denoted by colored lines as shown in Figs. \ref{fig_2018njp} (a) and (b) for S\&P 500 and Nikkei 225, respectively, as a function of the number of clusters.  Figs. \ref{fig_2018njp} (c) and (d) show the transition counts of paired market states for S\&P 500 and Nikkei 225 markets, respectively. The optimum number of cluster for \textit{$k\geq 4$} is chosen such as the standard deviation of the ensemble is lowest keeping the number of clusters highest. Our analysis depicts that $k=4$ and $5$ are the optimal number of clusters for S\&P 500 and Nikkei 225, respectively. We cluster similar correlation matrices into these optimized $k$ number of ``market states''. The market evolves through the transitions between these market states. Sometimes the market remains in a particular state for a longer time and sometimes it jumps shortly to another state and bounces back or evolves further. Transitions to nearby states are highly probable. Highest transition counts of market states occurs when market remains in the same state i.e. along the diagonal followed by first off diagonal as transitions to nearby states. The jump from the lowest $S1$ and $S2$ state to the critical state ($S4$ in USA and $S5$ in JPN) is less probable which acts as a precursor to the critical state and as a best state to hedge against the critical state.

Recently, Pharasi et. al. in Ref. \cite{pharasi2020dynamics} have presented an improved criteria for the classification of ``market state”, mainly due to the increased attention to the transition matrix of paired states. The optimization of the number of market states is done by using the following two parameters- number of clusters and noise suppression. However, here the preference is given to the transition matrix which avoids large jumps from lower states to the critical state. It brings a better perspective to hedge against the market’s critical events without a prohibitive cost. By using the same overlapping epoch of length 20 days but the shifts of 1 day, the results showed significantly improved statistics as compared to the shift of 10 days in Ref\cite{Pharasi_2018}. Figure \ref{fig_MS2020_epsVsk} (a) and (b) show the identification of number of market states for S\&P 500 ($N=350$ stocks) and Nikkei 225 ($N=156$ stocks) markets, respectively, over 14-year period from 2006-2019 using overlapping epochs of length $T=20$ days with shift of $\Delta=1$ day. In contrast to Fig. \ref{fig_2018njp}, here we use ensemble of 1000 different initial conditions for different number of clusters $k$ and noise suppression parameter $\epsilon$. {It is found that the minimum of $\sigma_{d_{intra}}$ for $k\geq 4$ at $k=5$ and $\epsilon=0.9$ for S\&P 500 and at $k=7$ and $\epsilon=0$ for Nikkei 225 markets}. This optimum number of cluster is used for the k-means clustering for 3503 correlation frames with noise-suppression $(\epsilon=0.9$) for S\&P 500 market and 3439 correlation frames without noise-suppression ($\epsilon=0$) for Nikkei 225 market. The 2D projection of 3D $k$-means clustering for S\&P market on $xz$-plane with $k=5$ clusters and Nikkei 225 market on $xy$-plane with $k=7$ clusters, respectively, are shown in Fig. \ref{fig_MS2020_epsVsk} (c) and (d). The distribution of clusters is different for two market- for S\&P 500 market, $k$-means clustering divides the distribution of correlation frames in 3D correlation space in the horizontal direction ($S1, S2,S3, S4, S5$) but for the Nikkei 225, it divides in both vertically stacked (S2 \& S3 and S4 \& S5) as well as horizontal directions.

Figure \ref{fig_avg_corr_mat_MS} (a) and (b) depict the schematic diagram of average correlation matrices of each market state for S\&P 500 and Nikkei 225 markets, respectively. From lower (calm period) to higher (crisis period) market states, mean correlation as the correlation of intra- and inter-sectorial correlation increases.  Figure \ref{fig_avg_corr_mat_MS} (c) and (d) show the color map plots of the transition counts of paired market states (MS) for S\&P 500 with $k=5$ clusters and noise-suppression parameter $\epsilon=0.9$ and Nikkei 225 market with $k=7$ clusters and without noise-suppressed ($\epsilon=0$), respectively over the period of 2006-2019. Transition matrix for S\&P 500 market  is nearly tridiagonal and shows zero transitions from lower states (S1, S2, and S3) to critical state (S5) with twelve transitions from penultimate state (S4) to critical state (S5). Thus, S4 behaves as a precursor to the market crashes and a good agent to hedge. For Nikkei 225 market, the transition counts of the paired market states between S2 and S3 (S2$\rightarrow$ S3=25, vertically stacked) are smaller than S2 and S4 (S2$\rightarrow$S4=45, in horizontal direction). The same behavior is noticed during the transitions between states S4 and S5, and S4 and S6. Therefore, state S6 behaves as a precursor to the critical state S7 with one transition, contrary to S\&P 500, from S5 to critical state S7. 

\section{Sectorial analysis}
\begin{figure}[hbt!]
    \centering
{\includegraphics[width=1\textwidth]{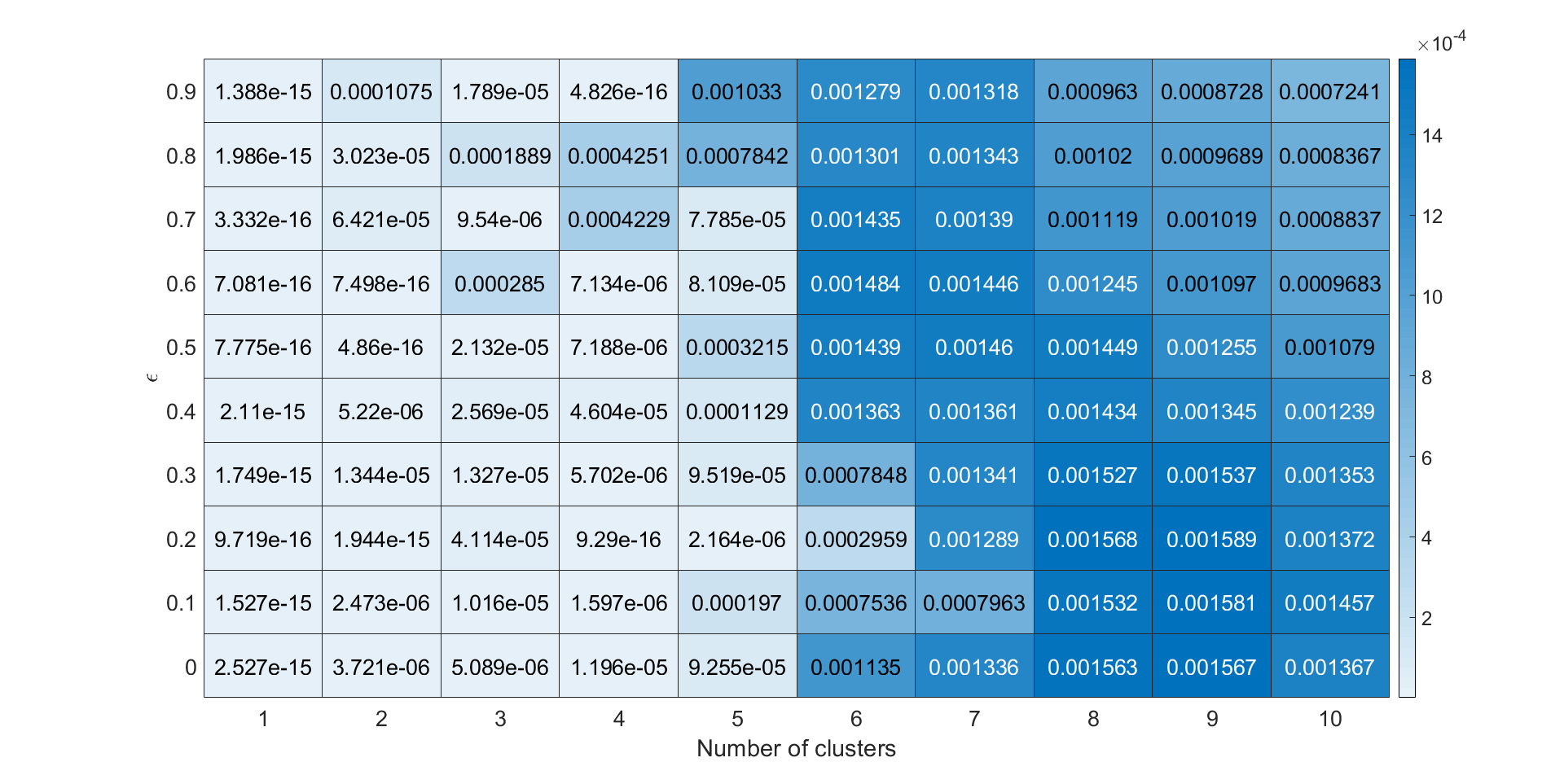}}\llap{\parbox[b]{6.6in}{\textbf{{\Large (a)}}\\\rule{0ex}{3.4in}}}
{\includegraphics[width=1\textwidth]{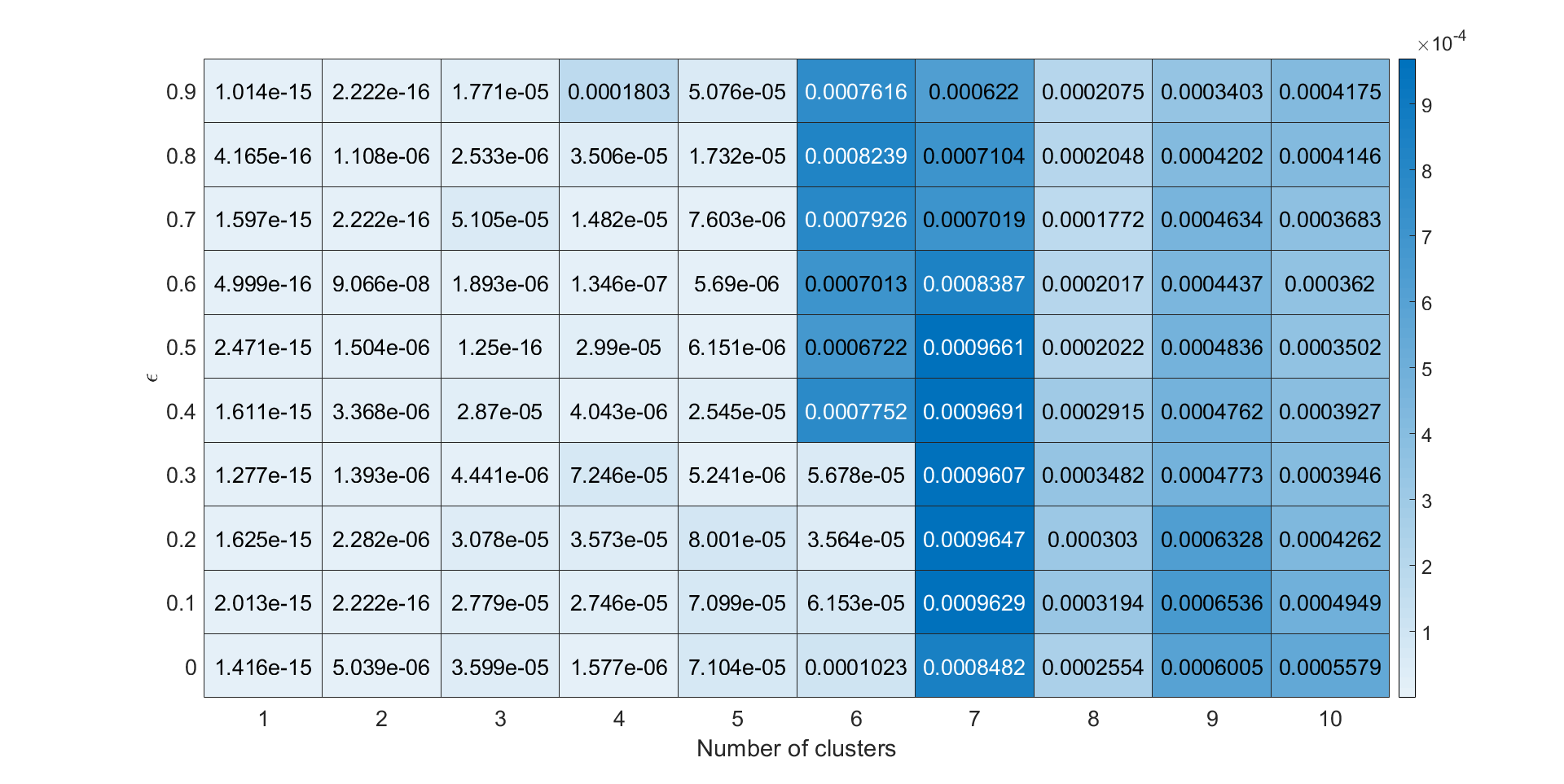}}\llap{\parbox[b]{6.6in}{\textbf{{\Large (b)}}\\\rule{0ex}{3.4in}}}
    \caption{Optimization of number of market states based on the minimum standard deviation of intra cluster distances ${\sigma_{d_{intra}}}$ for (a) S\&P 500, and (b) Nikkei 225 markets. We use MDS map, constructed from noise suppressed sectorial averaged matrices $M$, for $k$-means clustering. We use 1000 different initial conditions for the calculation of ${\sigma_{d_{intra}}}$. According to the minimum of standard deviations of intra cluster distances, the best choice ($k>4$)  for  (a) S\&P 500 is $k=5$ clusters and $\epsilon=0.2$, and for Nikkei 225,  $k=5$ clusters and $\epsilon=0.3$.}\label{Avg_k_optimum}
\end{figure}

We consider here the same data of $N=350$ stocks of the S$\&$P 500 index ($N=156$ stocks of the Nikkei 225) over the period from 2006-2019 with $T_{tot}= 3523$ $(T_{tot}= 3459)$  trading days\cite{pharasi2020dynamics}. We then calculate a sectorial averaged matrices $M$ by taking intra- and inter- sectorial average of the correlation matrices $C$ constructed from the overlapping epoch of $T=20$ days and shifted by $\Delta=1$ day. The symmetric matrices $M$ are not correlation matrices any more and the diagonal elements are not trivial.
We adopt the similar approach used by Rinn et. al. \cite{rinn2015dynamics}, where they performed a cluster analysis of sectorially averaged correlation matrices and stochastic process analysis to identify financial market dynamics by a few important variables. The market showed dynamical stability and transitions between the stable quasi-stationary states. The method allows to project the dynamics  of high-dimensional system to a low-dimensional space. In this manner, we have the matrices $M$ with dimension $N_{S}\times N_{S}$, where $N_{S}$ is the number of the sectors in the market- $N_{S}=10$ for S$\&$P 500 and $N_{S}=6$ for Nikkei 225. The number of independent variables in the $M$ matrix are $N_{S}(N_{S} +1)/2$.
The direct advantage of analyzing these matrices $M$, in comparison to full correlation matrices $C$, is their low dimensional space, i.e., 45D for S$\&$P 500  and 15D for Nikkei 225 markets which are much smaller than the original $N(N-1)/2$ dimensional space of correlation matrix $C$. Working in lower dimensional space has an advantage of reduced computational time and provides a useful tool for studying sectoral behavior.   
%On the other hand, comparing the result of sectoral and stocks analysis will give us better perspective on the crash risk protection.\\
We then apply the $k$-means clustering analysis on these matrices $M$ after applying the appropriate noise suppression \cite{guhr2003,vinayakpre2013}, based on the optimization using Fig. \ref{Avg_k_optimum}. We can achieve the same dimensional reduction by taking the  average of log-return time series for each sectors first and then calculate the correlations among these $N_{S}$ time series. But, we have found that this method is not effective as by taking average of time series, we average out a lot of information and unusual land up on high correlations even in the normal calm periods.
\begin{figure}[hbt!]
    \centering
    {\includegraphics[width=0.48\textwidth]{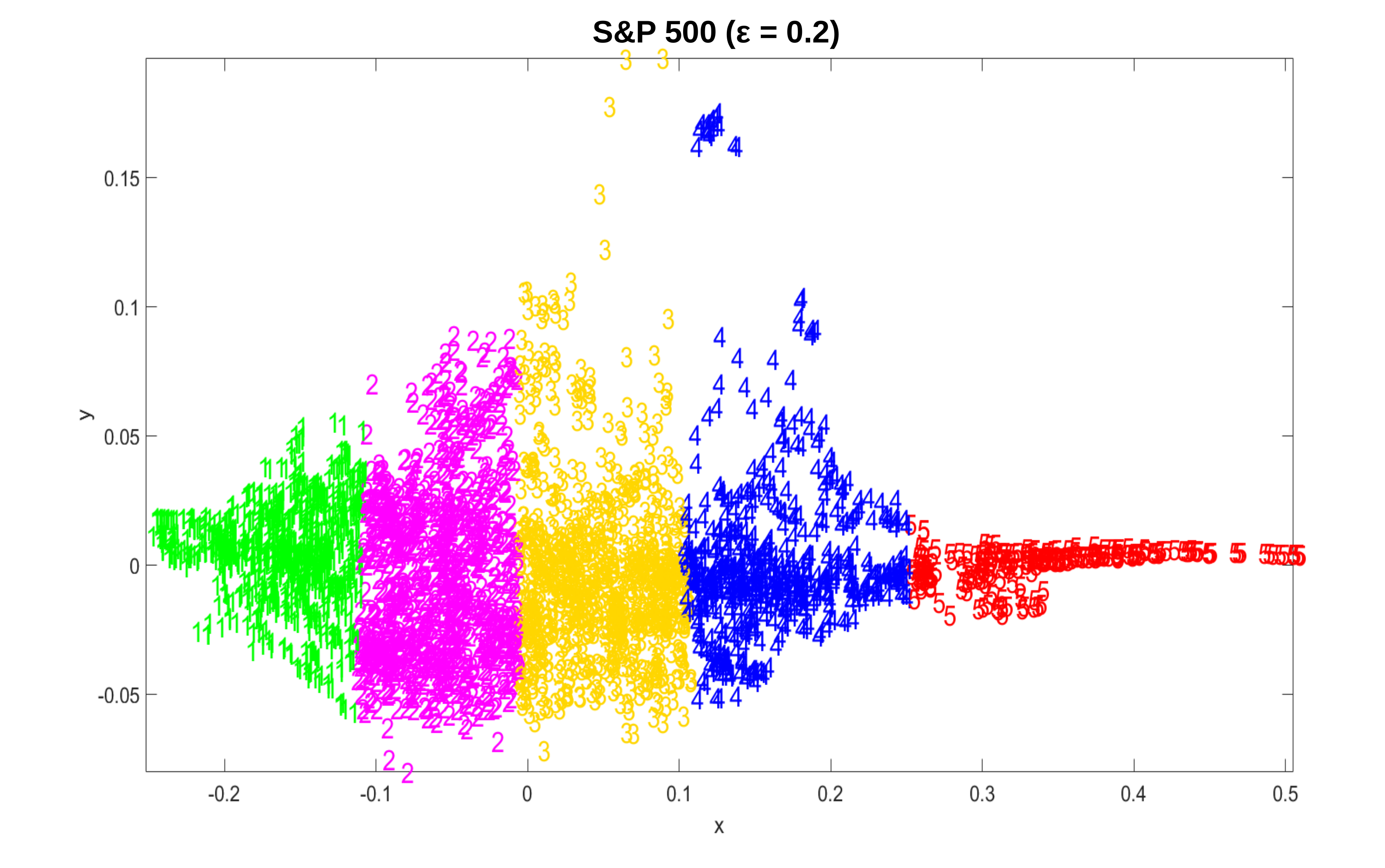}}\llap{\parbox[b]{3.2in}{\textbf{{\Large (a)}}\\\rule{0ex}{1.99in}}}
    {\includegraphics[width=.48\textwidth]{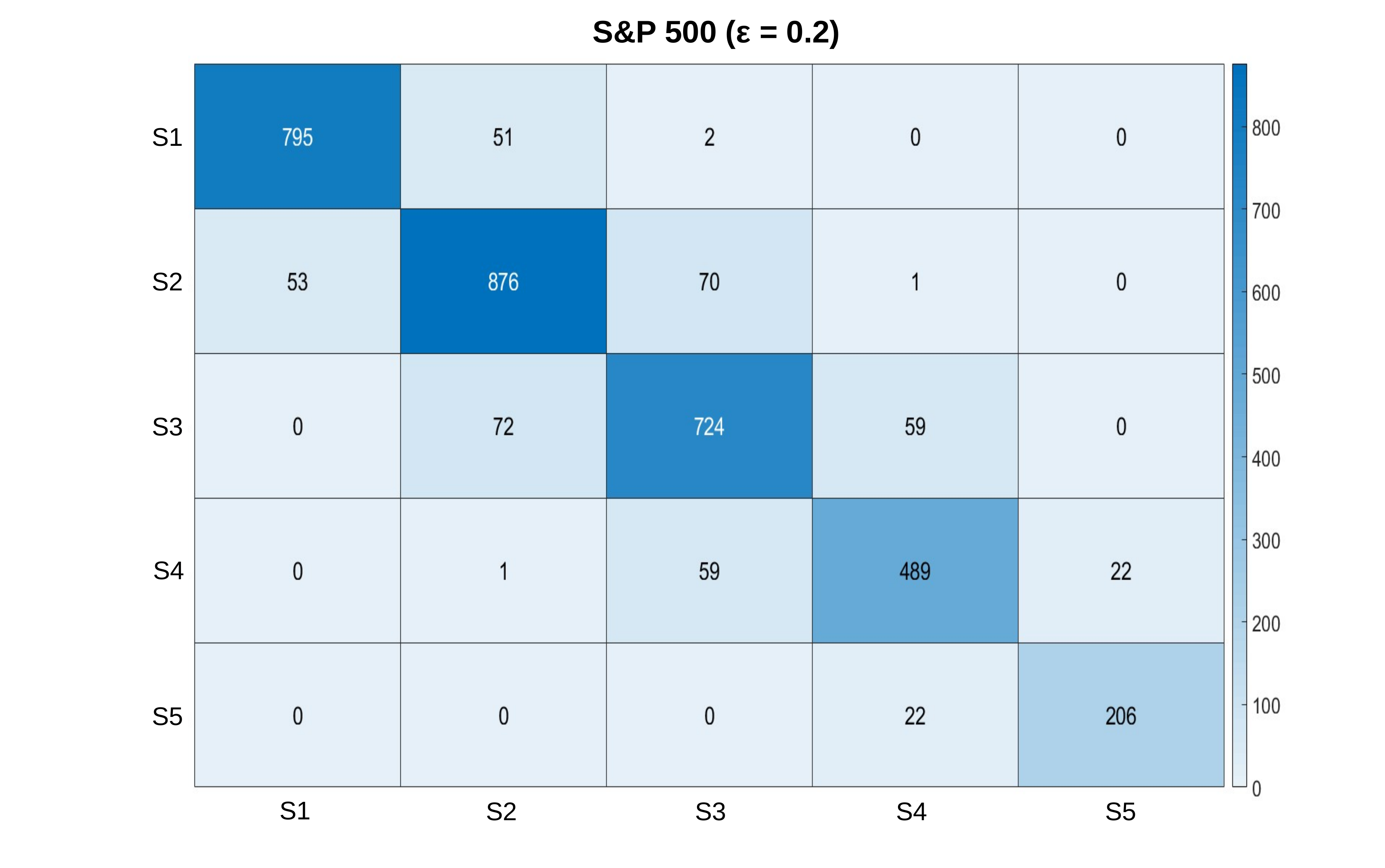}}\llap{\parbox[b]{3.2in}{\textbf{{\Large (b)}}\\\rule{0ex}{1.99in}}}
{\includegraphics[width=0.48\textwidth]{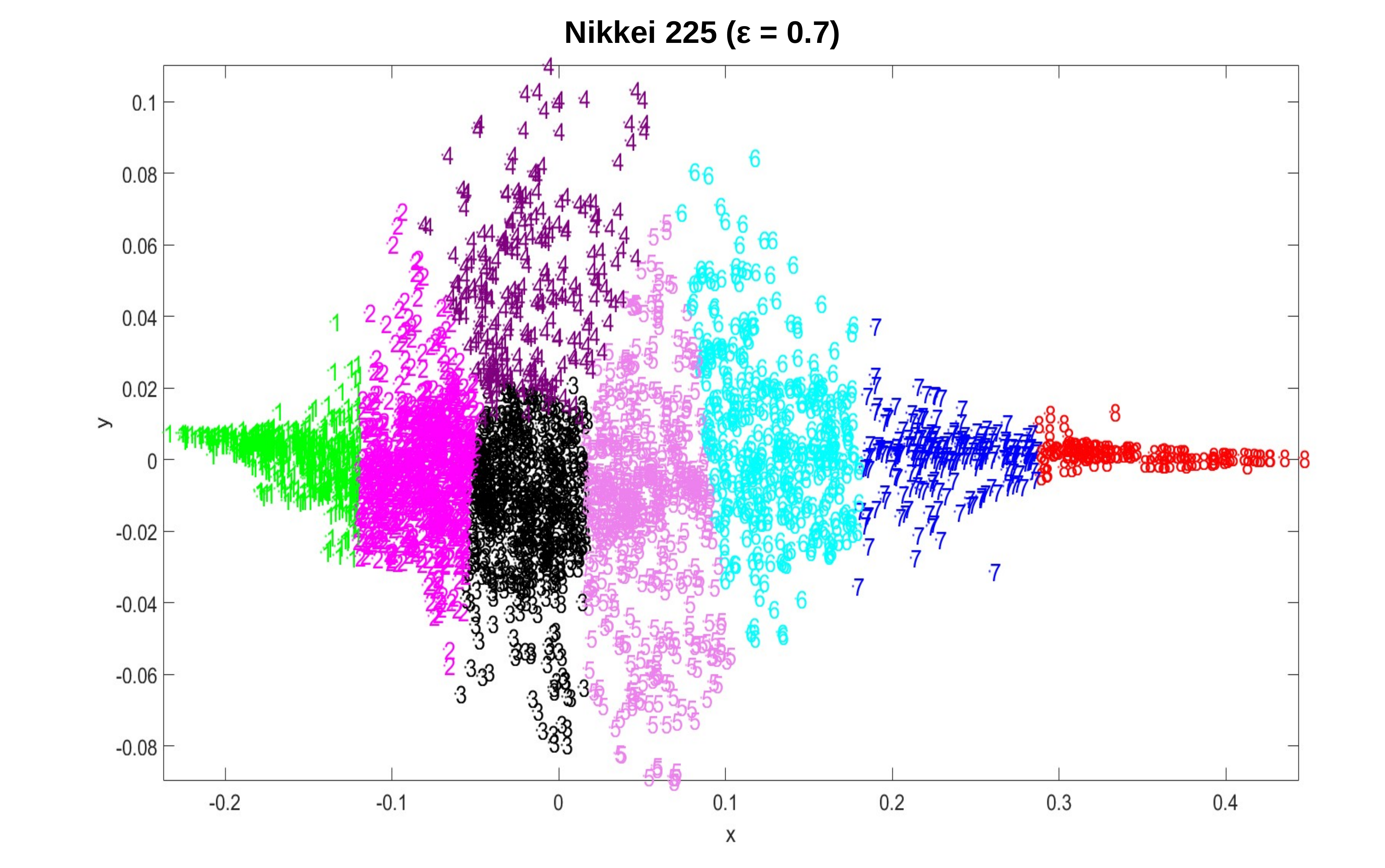}}\llap{\parbox[b]{3.2in}{\textbf{{\Large (c)}}\\\rule{0ex}{1.99in}}}
{\includegraphics[width=.48\textwidth]{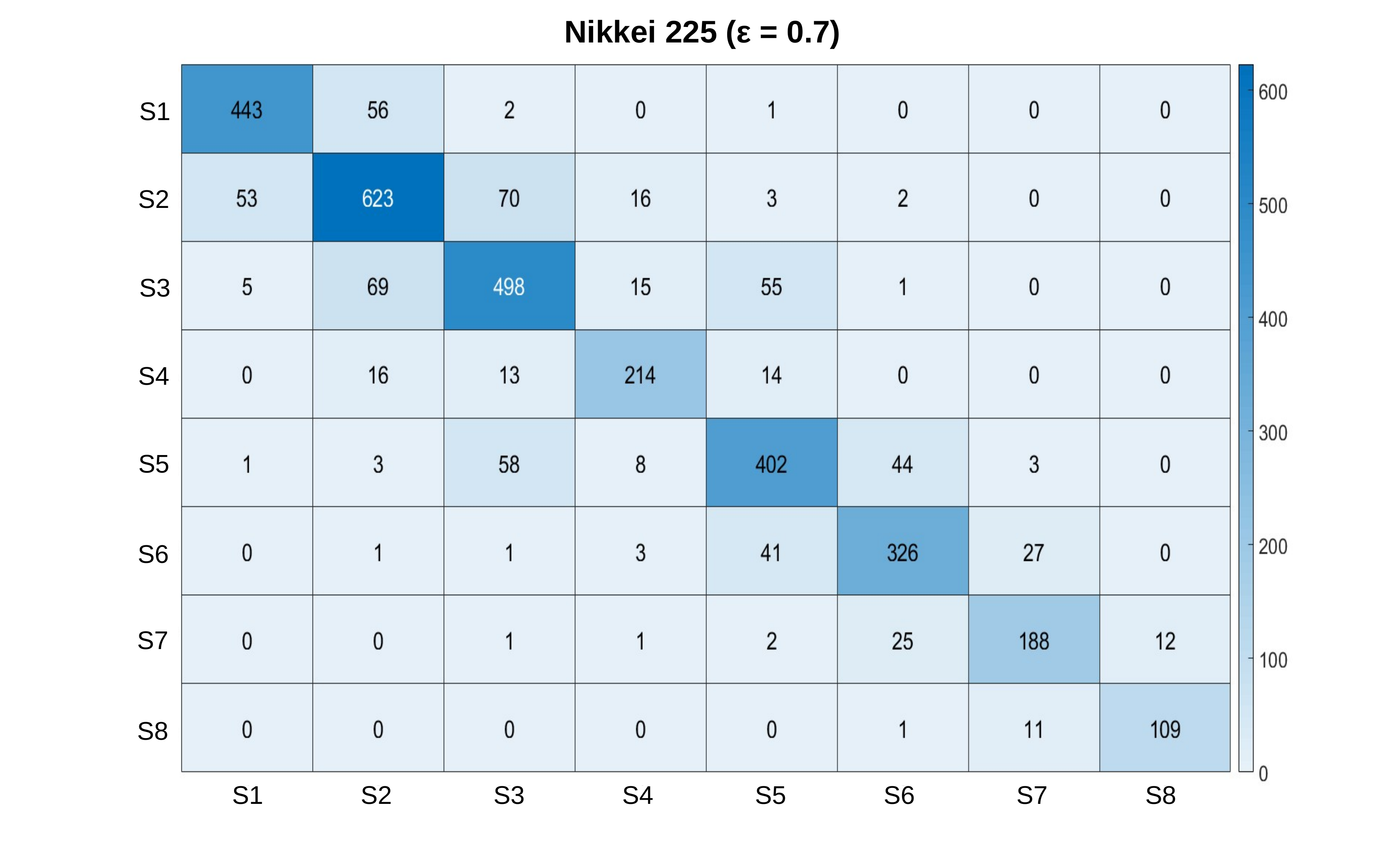}}\llap{\parbox[b]{3.2in}{\textbf{{\Large (d)}}\\\rule{0ex}{1.99in}}}
    \caption{Classification of the US market into five market states (top row) and the Japanese market into eight market states (bottom row). Plots show the 2D projection of 3D $k$-means clustering constructed from  noise suppressed matrices $M$ with (a) $\epsilon=0.2$ for S\&P 500 and (c) $\epsilon=0.7$ for Nikkei 225 markets. Color map plots for (b) S\&P 500 and (d) Nikkei 225 markets show the transition counts  of paired market states. For S\&P 500 market, we optimize the number of clusters using ${\sigma_{d_{intra}}}$ using Fig \ref{Avg_k_optimum} (a).  For the classification of the Japanese market into market states, we preferred eight clusters over five due to the transition matrix which avoids large jumps from lower states to the critical state.}\label{fig:sectorAvg}
\end{figure}

In Fig. \ref{Avg_k_optimum}, we optimize of number of market states based on the minimum standard deviation of intra cluster distances ${\sigma_{d_{intra}}}$. Fig. \ref{Avg_k_optimum} (a) and (b) show the optimization using intra- and inter- sectorial averaged matrices $M$ for S\&P 500 and Nikkei 225 markets, respectively. We  apply $k$-means clustering on MDS map, constructed from noise suppressed sectorial averaged matrices $M$. We use 1000 different initial conditions for the calculation of ${\sigma_{d_{intra}}}$. According to the minimum of standard deviations of intra cluster distances, the best choice ($k>4$)  for S\&P 500 is $k=5$ clusters and $\epsilon=0.2$ (see, Fig. \ref{Avg_k_optimum} (a)), and for Nikkei 225,  $k=5$ clusters and $\epsilon=0.3$ (see, Fig. \ref{Avg_k_optimum} (b)).

Figure \ref{fig:sectorAvg} shows the classification of the sectorial averaged matrices $M$ of S\&P 500 and Nikkei 225 markets. The US market is clustered into five market states and corresponding $k$-means clustering and transition matrix of paired market states are shown in Fig. \ref{fig:sectorAvg} (a) and (b), respectively. Here we choose $k=5$ clusters based on the robust measure (minimum of ${\sigma_{d_{intra}}}$ for $k>4$) of $k$-means clustering using Fig. \ref{Avg_k_optimum} (a). We have found that the $k$-means clustering and transition matrix show similar behavior as shown in the sectorial analysis (see, Fig. \ref{fig_MS2020_epsVsk}). But for the Japanese market, we preferred eight clusters over five due to ($i$) the transition matrix which avoids large jumps from lower states to the critical state and ($ii$) the similar distribution of clusters as obtained from the stocks analysis (see, Fig. \ref{fig_MS2020_epsVsk} (d)). The Japanese market is more complex than USA market and the distribution of $k$-means clusters ($k=8$) is in both vertical and horizontal directions in JPN market which is only horizontal direction for USA market. Transition matrices, consists of transition counts between different states, for S\&P 500 and Nikkei 225 represent a valuable tool to analyze the behavior of critical state (S5) are shown in Figs. \ref{fig:sectorAvg} (b) and (d), respectively. In this method, we have found that the number of epochs in the crash state of S\&P 500 market are higher (false positive) than one obtained from stock analysis in Fig. \ref{fig_avg_corr_mat_MS}. Furthermore, by increasing the number of clusters provide us appropriate way in which risk assessment can be studied with further details. The  transition matrix, shown in Figs. \ref{fig:sectorAvg} (d) for Nikkei 225, with eight number of clusters has less number of epochs in critical state which would be helpful for better risk assessment. 

The displacement between stocks  analysis and  sector analysis for S\&P 500 with five market states and noise suppressed $\epsilon=0.2$ is equal to $465$ days. We must note that the maximum of displacement, in this case, is $d_i=\pm 1$, with respect to the stock analysis of the Ref.\cite{pharasi2020dynamics}. There are $3038$ epochs which remain in the same states and we have $309$ epochs with displacement of $d_i=-1$ and $156$ epochs with displacement of $d_i=+1$. {On the other hand, the maximum of displacement  for Nikkei 225 with eight market states and noise suppressed $\epsilon=0.7$ is equal to $\pm3$. We have $14,25,813,2074,419,94,0$ epochs  with displacement of $d_i=-3,-2,-1,0,1,2,3$, respectively.}

\section{Trajectories in the correlation matrix space}
In this section, we compare the COVID-19 crash of S\&P 500 and Nikkei 225 market with other critical events as well as normal periods of both markets. For this purpose, we  considered the daily adjusted closure price data and the observation period is given in Tables \ref{table:ellipsoid_usa} and \ref{table:ellipsoid_jpn}  ($N=475$ for S\&P 500 and $N=204$ for Nikkei 225 from November 2019-June 2020). We consider the observation period of $125$ trading days keeping the crash day at the center.
%%%%%%%%%%%%%%%%%%%%%%%================================

\begin{table}[b!]	
\begin{center}
\caption{The table shows the correlation, measured for all consecutive pairs of co-ordinates of MDS at various dimensions. Here,
 $D_1=1$, $D_2=2$, $D_3=3$, $D_4=4$, $D_{\text{max}}=$size of $\zeta$. We consider here crashes in $2008$, $2010$, $2011$, $2015$, COVID-$19$, $1987$, 
two normal periods- $2006$ and $2017$, and for the $14$ years' data from $2006-2019$ for the S\&P 500 market. Except for the last case, we have taken the data of $125$ trading days.
 The correlation is more between a higher dimension with the maximum dimension (size of $\zeta$).}\label{table:corr_euclid_dist}
\begin{tabular} { | p {2 cm} | p {1.1 cm} | p {1.1 cm} | p {1.1 cm} | p {1.1 cm} | p {1.1 cm} |p {1.1 cm} | p {1.1 cm} | p {1.3 cm} | p {1.1 cm} |}
\hline
Correlation between & 1987 & 2006 & 2008 & 2010 & 2011 & 2015 & 2017 & COVID-19 & 2006-2019\\[2ex]
\hline
D$_1$, D$_{\text{max}}$ & 0.9455 & 0.7360 & 0.7065 & 0.8817 & 0.8957 & 0.8940 &  0.7916 & 0.8602  & 0.8361\\[2ex] 
D$_2$, D$_{\text{max}}$ & 0.9490 & 0.7673 & 0.7384 & 0.9035 & 0.9205 & 0.9010 & 0.8797 & 0.8731 & 0.8689\\[2ex]
D$_3$, D$_{\text{max}}$ & 0.9569 & 0.8407 & 0.7776 & 0.9253 & 0.9362 & 0.9119  & 0.9146 & 0.9081 &0.8802\\[2ex]
D$_4$, D$_{\text{max}}$ & 0.9633 & 0.8561 & 0.8694 & 0.9374 & 0.9399 & 0.9279  & 0.9274 & 0.9145 & 0.8855\\[2ex]
\hline
\end{tabular}
\end{center}
\end{table}
%%%%%%%%%%%%%%%%%%%%%%%===============================
\begin{figure}[ht!]
\centering
\includegraphics[width=0.49\linewidth]{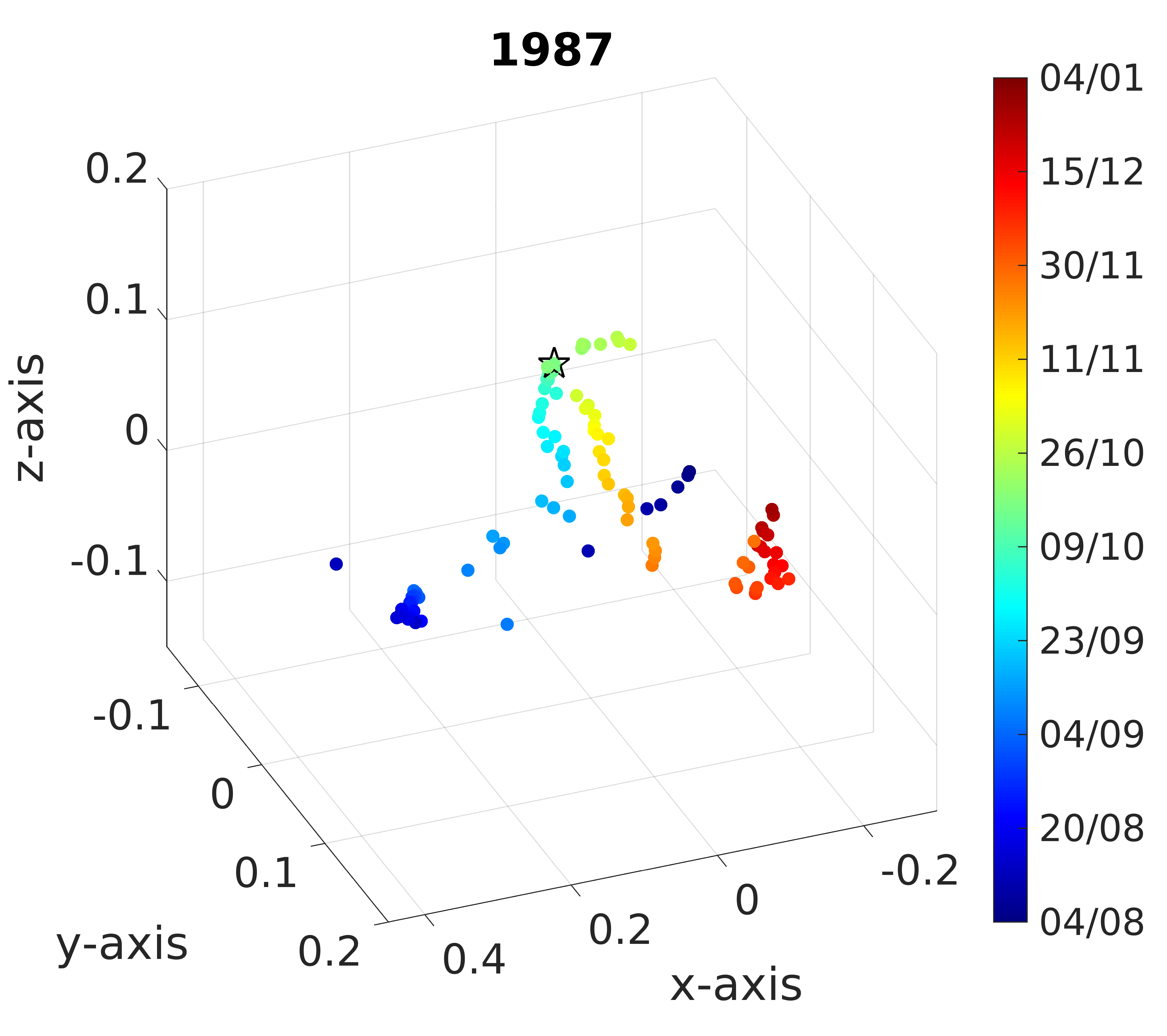}\llap{\parbox[b]{3.3in}{\textbf{\Large (a)}\\\rule{0ex}{2.9in}}}
\includegraphics[width=0.49\linewidth]{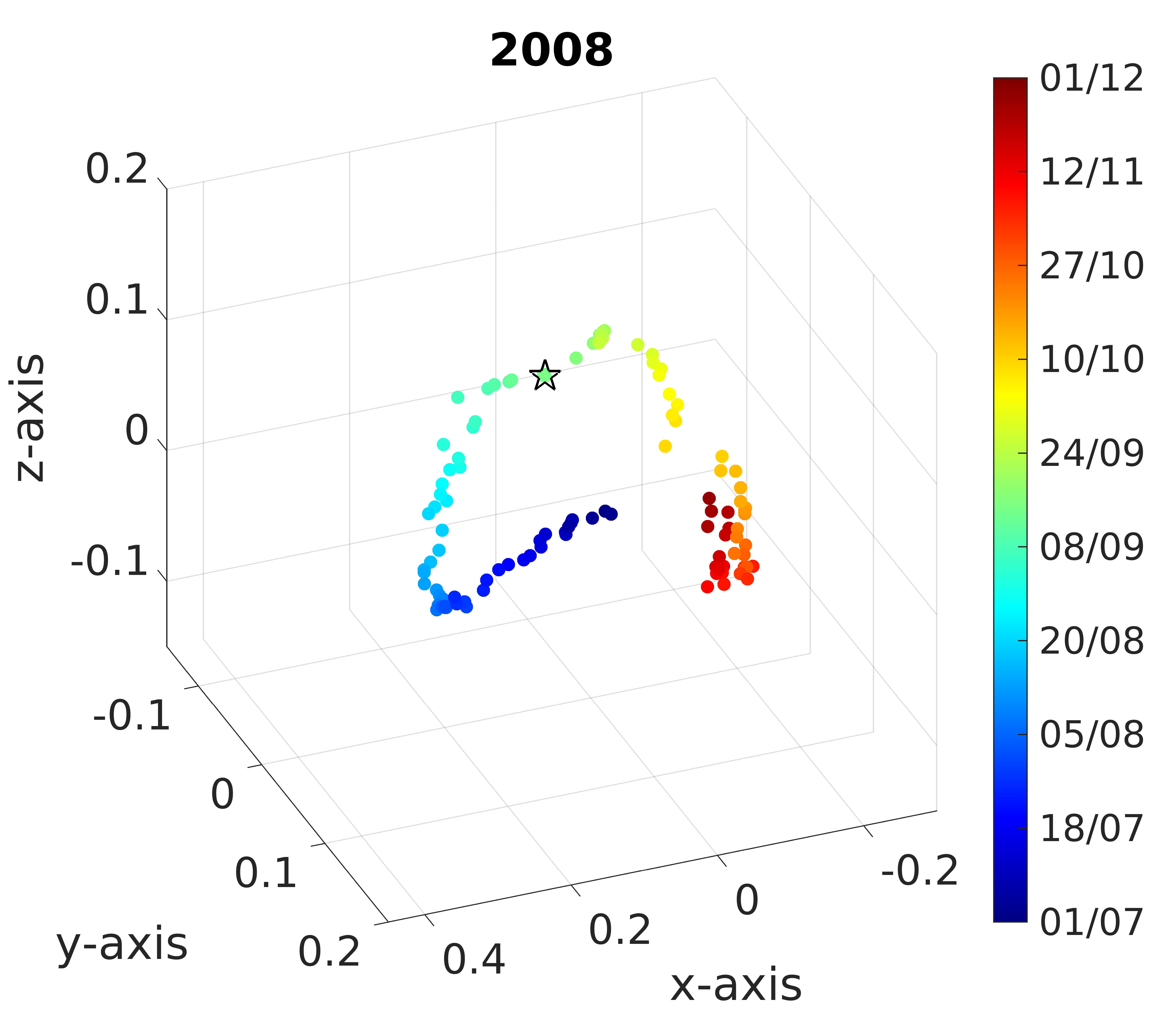}\llap{\parbox[b]{3.3in}{\textbf{\Large (b)}\\\rule{0ex}{2.9in}}}\\
\includegraphics[width=0.49\linewidth]{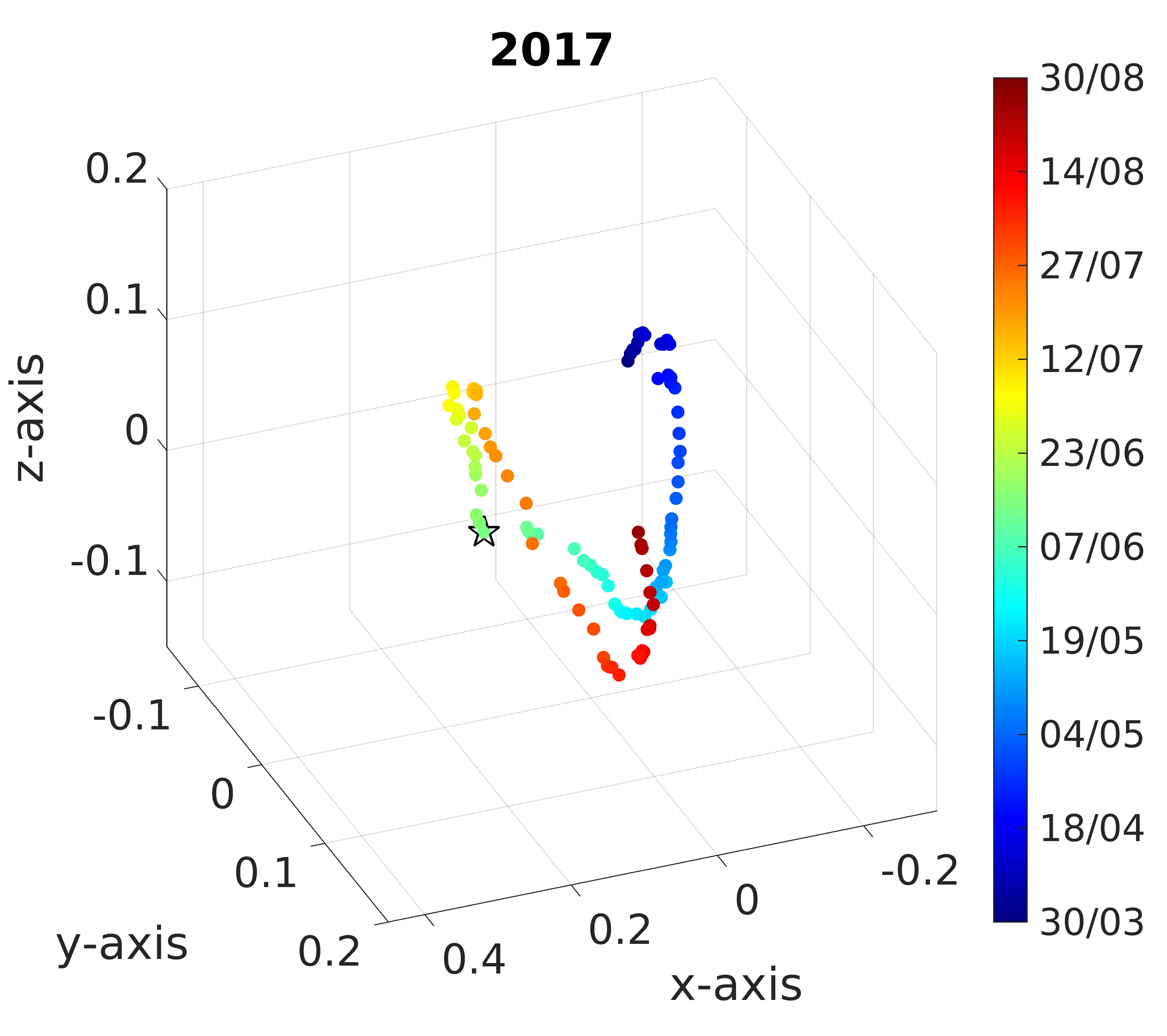}\llap{\parbox[b]{3.3in}{\textbf{\Large (c)}\\\rule{0ex}{2.9in}}}
\includegraphics[width=0.49\linewidth]{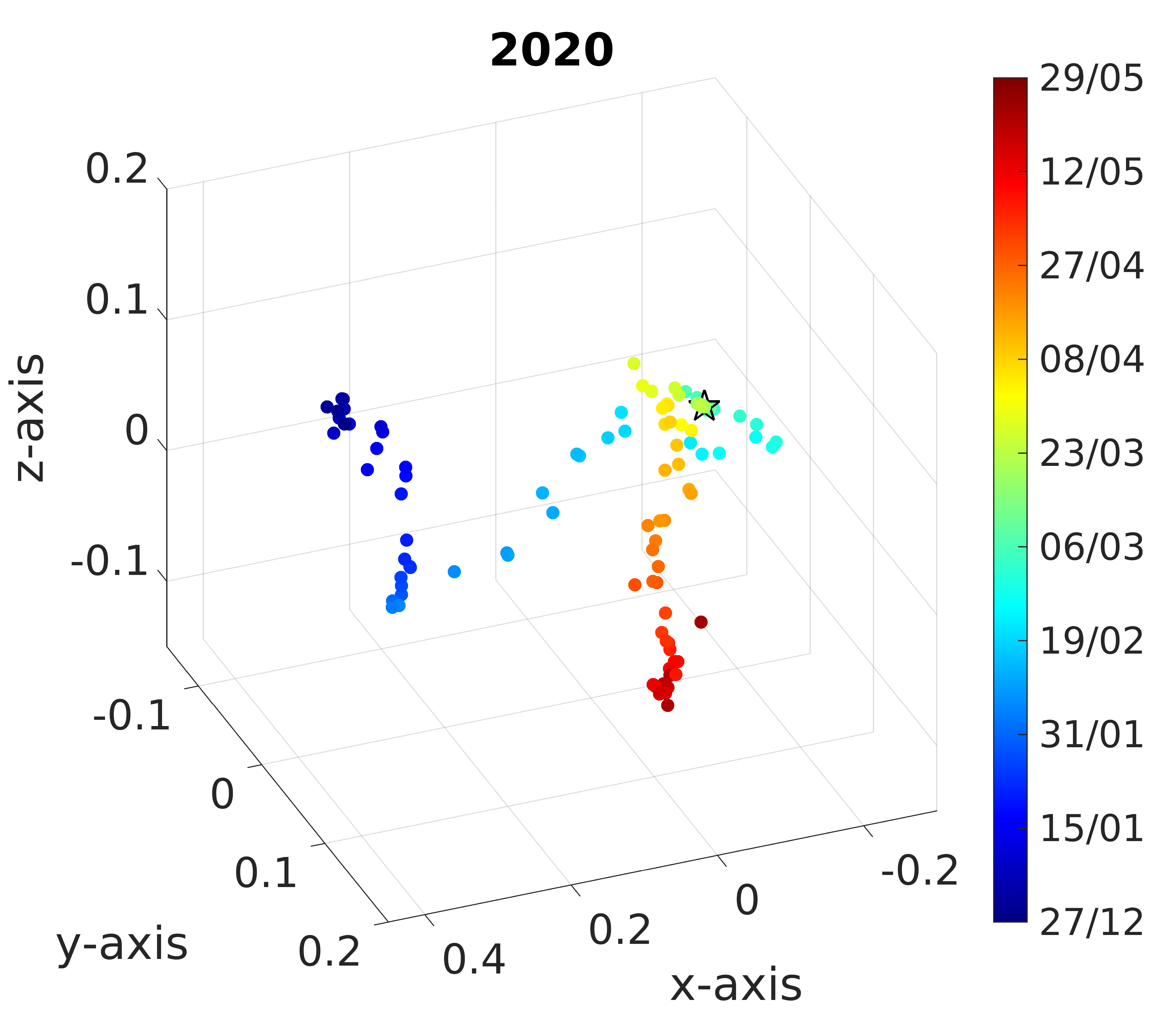}\llap{\parbox[b]{3.3in}{\textbf{\Large (d)}\\\rule{0ex}{2.9in}}}
\caption{Evolution of the trajectories of the (dis-)similarity measure between consecutive correlation matrices in the correlation matrix space for S\&P 500 market. The trajectories in 3D space is the projection of $Fr(Fr-1)/2$ dimensional space of the similarity matrix $\zeta$, with dimension $Fr\times Fr$, using MDS for S\&P 500 market for different time horizons: (a) $1987$ Black Monday crash, (b) 2008 Lehman Brother crash, (c) $2017$ normal period, and (d) COVID-19 crash. For the analysis, we have considered a period of $125$ trading days keeping the crash date at the center (hollow black star). The trajectory consists of equidistant points with some intermittent abrupt increments during critical events. The (dis-)similarity measure is the distance among correlation matrices so when the transition occurs between two consecutive correlation matrices in the same or nearby market states then the corresponding distance measure shows small and nearly equidistant increments. When the transitions happen between different market states then the corresponding distance measure shows big increments. We have also compared here three crashes and one normal period.}
\label{snake_usa}
\end{figure}

\begin{figure}[hbt!]
\centering
\includegraphics[width=0.49\linewidth]{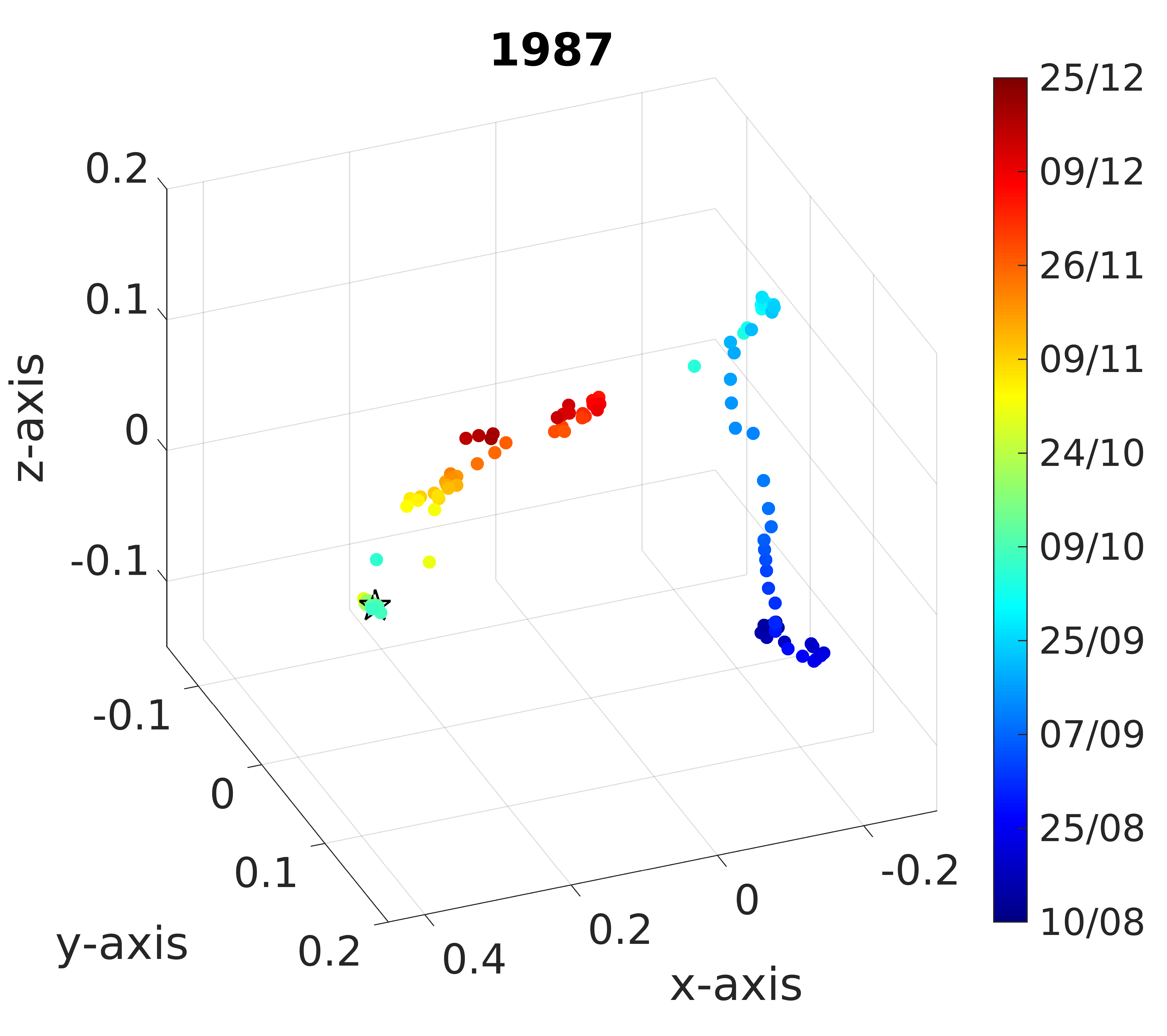}\llap{\parbox[b]{3.3in}{\textbf{\Large (a)}\\\rule{0ex}{2.9in}}}
\includegraphics[width=0.49\linewidth]{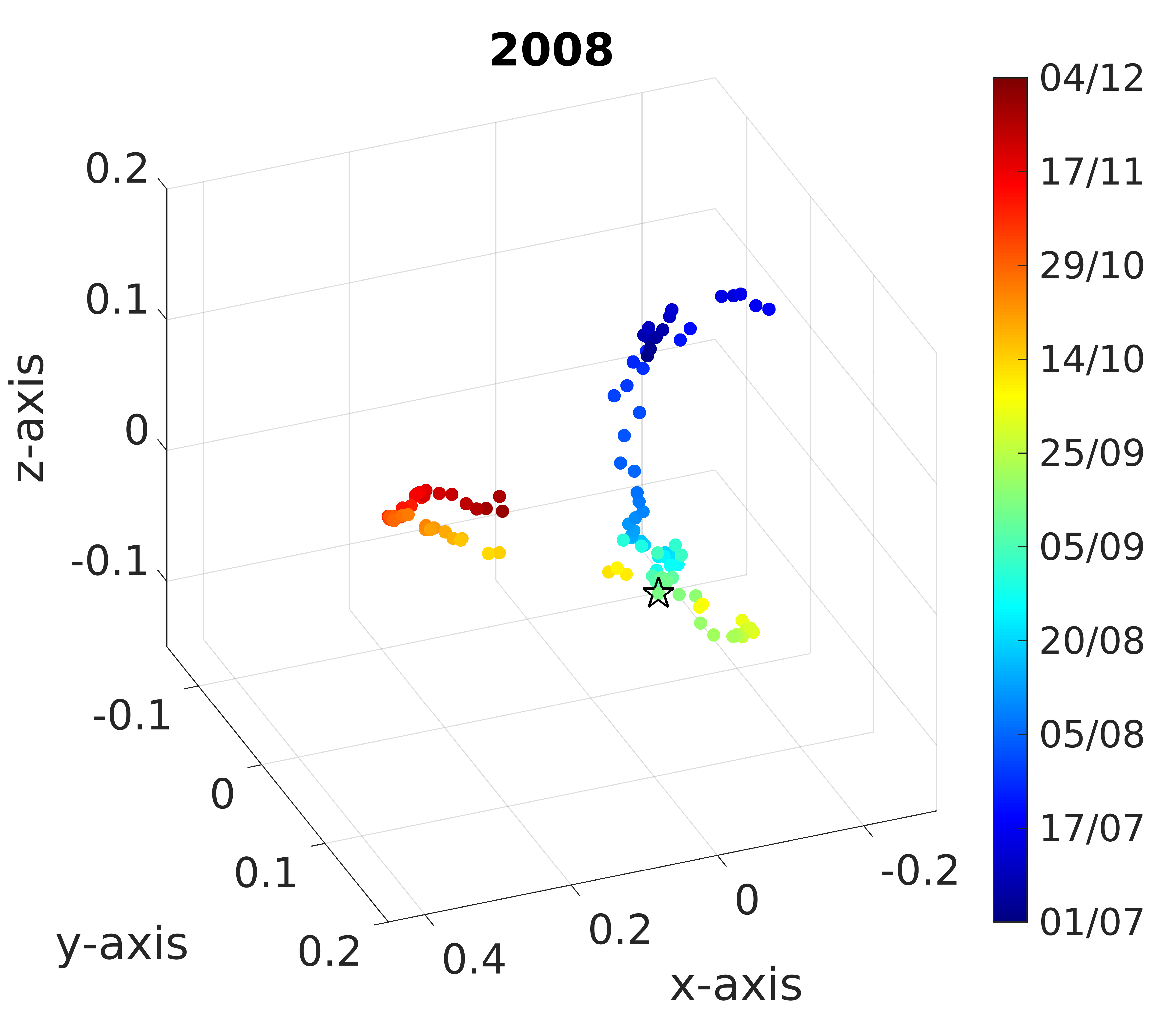}\llap{\parbox[b]{3.3in}{\textbf{\Large (b)}\\\rule{0ex}{2.9in}}}\\
\includegraphics[width=0.49\linewidth]{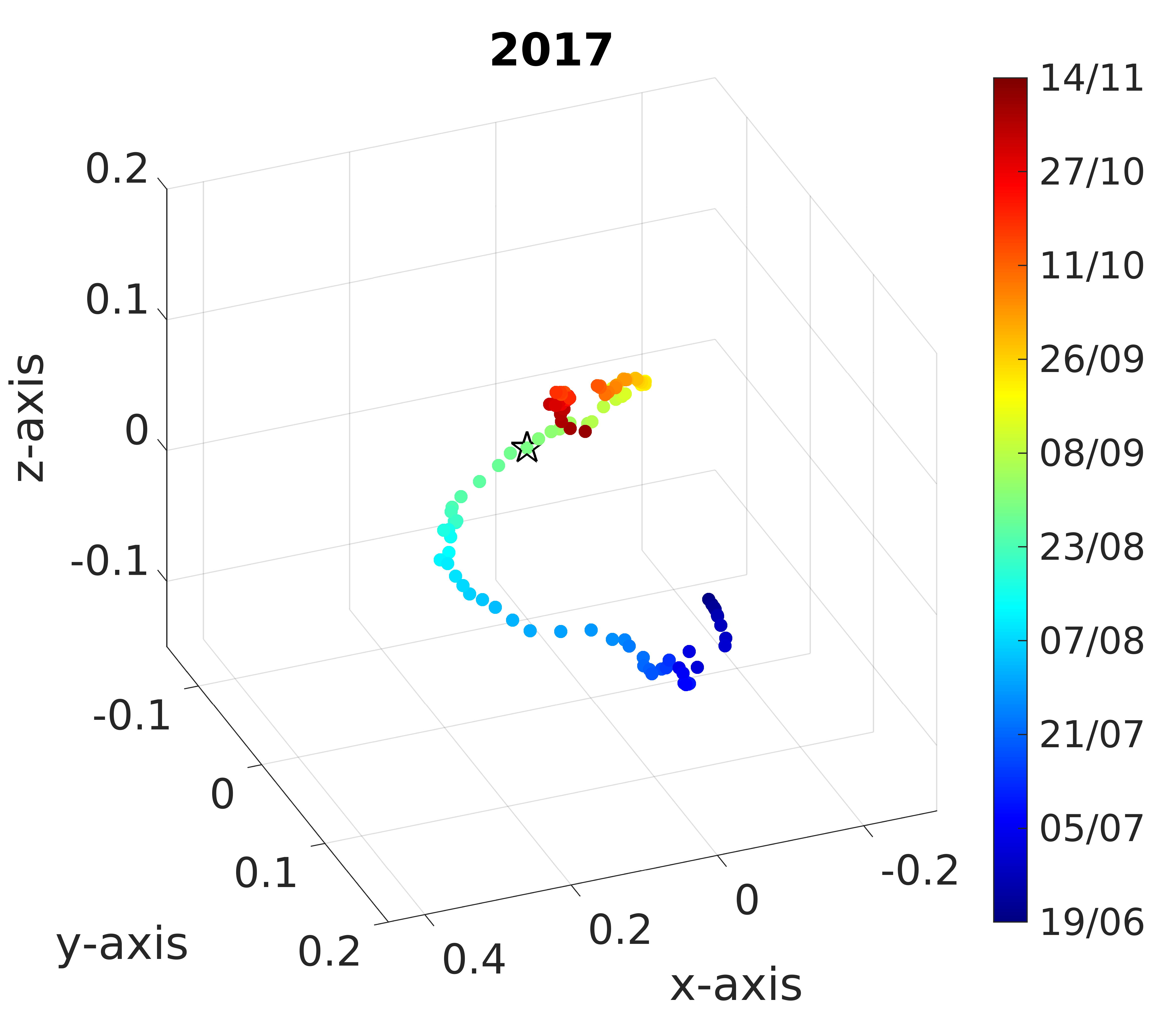}\llap{\parbox[b]{3.3in}{\textbf{\Large (c)}\\\rule{0ex}{2.9in}}}
\includegraphics[width=0.49\linewidth]{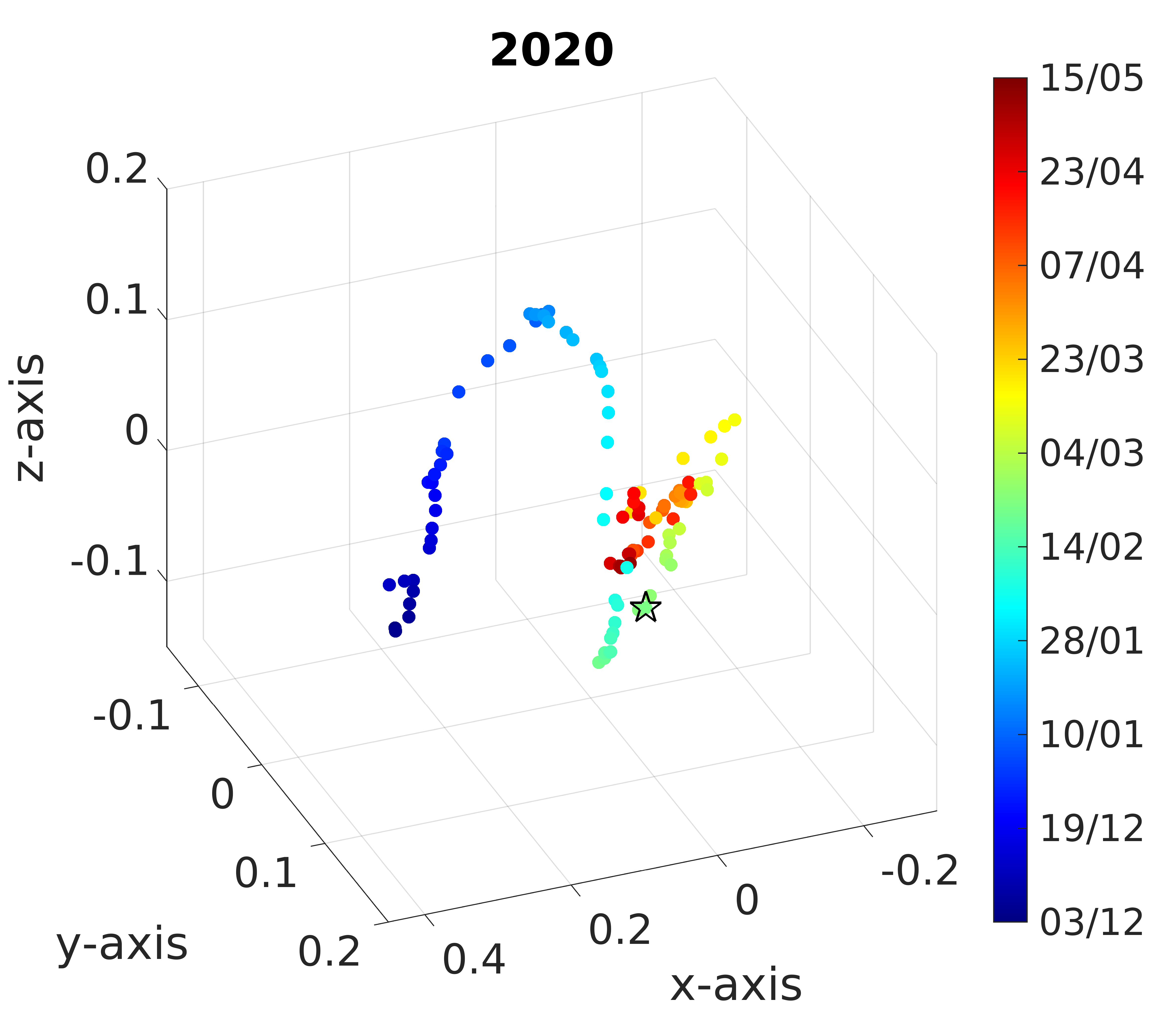}\llap{\parbox[b]{3.3in}{\textbf{\Large (d)}\\\rule{0ex}{2.9in}}}\\

\caption{Evolution of the trajectories of the (dis-)similarity measure between consecutive correlation matrices in the correlation matrix space for Nikkei 225 market. The trajectories in 3D space is the projection of $Fr(Fr-1)/2$ dimensional space of the similarity matrix $\zeta$, with dimension $Fr\times Fr$, using MDS for S\&P 500 market for different time horizons: (a) $1987$ Black Monday crash, (b) 2008 Lehman Brother crash, (c) $2017$ normal period, and (d) COVID-19 crash. For the analysis, we have considered time period of $125$ trading days keeping the crash date at the center (hollow black star). We have also compared here three crashes and one normal periods.}
\label{snake_jpn}
\end{figure}
%%%===================================================
\subsection{Comparison of COVID-19 case with other crash and normal periods}
We use MDS that reproduces the original metric or distances. Classical multidimensional scaling at various dimensions (D=$1, 2, 3, 4$, size of $\zeta$) of (dis)similarity matrix produces co-ordinates from a D-dimensional map. We can measure the Euclidean distance between pairs of all co-ordinates. Table {\ref{table:corr_euclid_dist}} shows the correlation between the Euclidean distances at various dimensions for various crashes and normal periods.  We consider here crashes at $2008$, $2010$, $2011$, $2015$, COVID-$19$, $1987$, two normal periods- $2006$ and $2017$, and for the $14$ years data from $2006-2019$. Except for the last case, we have taken the data of $125$ trading days. It is evident that the correlation coefficient is maximum for $4$D and the size of $\zeta$ for all the analysis periods.

Figure \ref{snake_usa}  and \ref{snake_jpn} show the dynamics of the (dis-)similarity measure between consecutive correlation matrices in the correlation matrix space for S\&P 500 and Nikkei 225 markets, respectively. In the figures, we use MDS map to show the trajectories in 3D space for different time horizons: (a) $1987$ Black Monday crash, (b) 2008 Lehman Brother crash, (c) $2017$ normal period, and (d) COVID-19 crash. The crash date is highlighted with hollow black star symbol in the figure. Each point in the figure corresponds to a correlation matrix of 20 days epoch. The distance among correlation matrices is small when market remains in the same state or transition occurs in nearby market states showing nearly equidistant increments in the trajectory whereas when the transitions happens from lower states to higher market states during critical events then corresponding distance measure shows some intermittent abrupt increments.
%\newpage
\begin{table}[htb!]
    \centering
    \begin{tabular}{|l|l|l|l|l|l|l|l|}
    \hline
        Serial No & Period & Starting date & End date & ${\sigma_x}^2$ & ${\sigma_y}^2$ & ${\sigma_z}^2$ & ${\sigma_r}^2$ \\ \hline
        1 & Black Monday crash & Aug 4, 1987 & Jan 5, 1988 & 0.0248 & 0.0035 & 0.0029 & 0.142 \\ \hline
        2 & August 2011 fall & May 23, 2011 & Oct 21, 2011 & 0.0213 & 0.0029 & 0.0017 & 0.1349 \\ \hline
        3 & DJ Flash crash & Feb 19, 2010 & Jul 22, 2010 & 0.034 & 0.0038 & 0.0022 & 0.1109 \\ \hline
        4 & Lehman Brothers crash & Jul 1, 2008 & Dec 1, 2008 & 0.0239 & 0.0035 & 0.0026 & 0.1442 \\ \hline
        5 & Covid-19 crash & Dec 27, 2019 & Jun 1, 2020 & 0.0303 & 0.0062 & 0.0032 & 0.2059 \\ \hline
        6 & Brexit & Apr 8, 2016 & Sep 8, 2016 & 0.0212 & 0.0054 & 0.0038 & 0.2538 \\ \hline
        7 & IPO Facebook Debut & Mar 5, 2012 & Aug 3, 2012 & 0.0089 & 0.0045 & 0.0034 & 0.5134 \\ \hline
        8 & Flash Freeze & Jun 7, 2013 & Nov 6, 2013 & 0.0114 & 0.0042 & 0.004 & 0.3661 \\ \hline
        9 & Treasury Freeze & Jul 31, 2014 & Dec 31, 2014 & 0.0156 & 0.0051 & 0.0038 & 0.3249 \\ \hline
        10 & Chinese Black Monday & Jun 9, 2015 & Nov 6, 2015 & 0.0255 & 0.0043 & 0.004 & 0.1671 \\ \hline
        11 & Normal-1(S1 to S2) & Jul 26, 2006 & Dec 26, 2006 & 0.0071 & 0.005 & 0.0046 & 0.7098 \\ \hline
        12 & Normal-1(S1 to S2) & Sep 12, 2006 & Feb 14, 2007 & 0.0061 & 0.0049 & 0.0047 & 0.8122 \\ \hline
        13 & Normal-1(S1 to S2) & Oct 10, 2016 & Mar 14, 2017 & 0.0081 & 0.0063 & 0.0048 & 0.7823 \\ \hline
        14 & Normal-1(S1 to S2) & Mar 30, 2017 & Aug 30, 2017 & 0.0074 & 0.0056 & 0.0041 & 0.7628 \\ \hline
        15 & Normal-1(S1 to S2) & Sep 1, 2017 & Feb 5, 2018 & 0.0079 & 0.0065 & 0.0051 & 0.8255 \\ \hline
        16 & Normal-2(S1 to S3) & Jun 1, 2006 & Oct 31, 2006 & 0.0124 & 0.0052 & 0.0039 & 0.4161 \\ \hline
        17 & Normal-2(S1 to S3) & Mar 27, 2007 & Aug 27, 2007 & 0.0105 & 0.0043 & 0.0036 & 0.4117 \\ \hline
        18 & Normal-2(S1 to S3) & Apr 21, 2009 & Sep 21, 2009 & 0.0059 & 0.0049 & 0.0029 & 0.8198 \\ \hline
        19 & Normal-2(S1 to S3) & Sep 28, 2010 & Mar 1, 2011 & 0.0124 & 0.005 & 0.0042 & 0.40 \\ \hline
        20 & Normal-2(S1 to S3) & Jul 10, 2012 & Dec 11, 2012 & 0.007 & 0.0051 & 0.0039 & 0.7345 \\ \hline
        21 & Normal-2(S1 to S3) & Jan 20, 2015 & Jun 22, 2015 & 0.0073 & 0.0051 & 0.004 & 0.7066 \\ \hline
        22 & Normal-2(S1 to S3) & Apr 24, 2018 & Sep 24, 2018 & 0.0082 & 0.0058 & 0.0047 & 0.7054 \\ \hline
    \end{tabular}
\caption{The table shows characterization of critical and normal events based on the variance of MDS map of correlation matrix trajectories of $125$ days ($6$ months approx.) for the S\&P 500 market. The critical events are kept at the center of the trajectory, constructed from the overlapping epoch of $T=20$ days and shifts of $\Delta=1$ day. The second column provides the name of the various events and corresponding time periods are shown in third (Starting date) and fourth (End date) columns. Fifth, sixth, and seventh columns show the variances ${\sigma_x}^2, {\sigma_y}^2, {\sigma_z}^2$, respectively, of the 3D MDS map using the similarity matrices. For the analysis, we take the ratio of variances of y- and x- directions (${\sigma_r}^2={\sigma_y}^2/{\sigma_x}^2$) in eighth column. We have found that the ratio stays below 0.4 for critical events and goes above 0.4 for normal periods. There is a false positive event, the ratio for the critical event IPO Facebook Debut (Serial No 7) shows the value higher than 0.4, we are still investigating the reason behind it.}
\label{table:ellipsoid_usa}
\end{table}

The Tables \ref{table:ellipsoid_usa}  and \ref{table:ellipsoid_jpn} show the analysis of characterization of critical and normal events based on the variance of MDS map of correlation matrix trajectories of $125$ days ($6$ months approx.) for the S\&P 500 and Nikkei 225 markets, respectively. We have considered data from various critical and normal periods, details are given in  
the second column of the table with the corresponding starting and ending date in third and fourth columns, respectively. We categorized the normal events into two groups: Normal-1 period, where the trajectory is bounded between states S1 and S2, and Normal-2 period, where the trajectory is bounded among three states S1 to S3 based on the characterization shown in Figs. \ref{fig_MS2020_epsVsk} and \ref{fig_avg_corr_mat_MS}. We use the variances of the 3D MDS map to calculate the ratio of variances of y- and x- directions (${\sigma_r}^2={\sigma_y}^2/{\sigma_x}^2$). Note that, the ratio stays below 0.4 for critical events and goes above 0.4 for normal periods. There is a false positive event for the S\&P 500 market where the value for the critical event IPO Facebook Debut (Serial No 7) goes above 0.4. Also, for Nikkei 225 market the ratio of variances for the critical events in August 2011 fall (Serial No 2) and IPO Facebook Debut (Serial No 8) show the value higher than 0.4. We are still investigating the reason behind it.

\begin{table}[htb!]
    \centering
    \begin{tabular}{|l|l|l|l|l|l|l|l|}
    \hline
        Serial No & Period & Starting date & End date & ${\sigma_x}^2$ & ${\sigma_y}^2$ & ${\sigma_z}^2$ & ${\sigma_r}^2$ \\ \hline
        1 & Black Monday crash & Aug 10, 1987 & Dec 26, 1987 & 0.0479 & 0.0055 & 0.0043 & 0.115 \\ \hline
        2 & August 2011 fall & May 24, 2011 & Oct 25, 2011 & 0.0083 & 0.004 & 0.0025 & 0.4768 \\ \hline
        3 & DJ Flash crash & Feb 15, 2010 & Jul 22, 2010 & 0.0118 & 0.0033 & 0.0024 & 0.2807 \\ \hline
        4 & Lehman Brothers crash & Jul 1, 2008 & Dec 4, 2008 & 0.0234 & 0.0038 & 0.0035 & 0.1638 \\ \hline
        5 & Covid-19 crash & Dec 3, 2019 & May 15, 2020 & 0.014 & 0.0038 & 0.0035 & 0.2707 \\ \hline
        6 & Tsunami/Fukushima & Dec 21, 2010 & Jun 1, 2011 & 0.0227 & 0.0042 & 0.0032 & 0.1835 \\ \hline
        7 & Brexit & Apr 5, 2016 & Sep 8, 2016 & 0.0168 & 0.0052 & 0.002 & 0.3072 \\ \hline
        8 & IPO Facebook Debut & Feb 29, 2012 & Aug 2, 2012 & 0.0088 & 0.0041 & 0.0031 & 0.4676 \\ \hline
        9 & Flash Freeze & Jun 7, 2013 & Nov 11, 2013 & 0.0071 & 0.0025 & 0.002 & 0.3526 \\ \hline
        10 & Treasury Freeze & Jul 29, 2014 & Jan 6, 2015 & 0.0185 & 0.0043 & 0.0032 & 0.2346 \\ \hline
        11 & Normal-1(S1 to S3) & May 9, 2017 & Oct 4, 2017 & 0.0084 & 0.006 & 0.0039 & 0.7171 \\ \hline
        12 & Normal-1(S1 to S3) & Jun 15, 2017 & Nov 10, 2017 & 0.0075 & 0.0062 & 0.0053 & 0.8251 \\ \hline
        13 & Normal-2(S1 to S4) & Apr 19, 2017 & Sep 19, 2017 & 0.0104 & 0.006 & 0.0038 & 0.582 \\ \hline
        14 & Normal-2(S1 to S4) & Jun 19, 2017 & Nov 14, 2017 & 0.0072 & 0.0061 & 0.0053 & 0.845 \\ \hline
        15 & Normal-2(S1 to S4) & Aug 15, 2017 & Jan 10, 2018 & 0.01 & 0.0052 & 0.0045 & 0.523 \\ \hline
    \end{tabular}
\caption{The table shows characterization of critical and normal events based on the variance of MDS map of correlation matrix trajectories of $125$ days ($6$ months approx.) for the Nikkei 225 market. The details of the method is given in Table \ref{table:ellipsoid_usa}. The second column provides the name of the various events and corresponding time periods are shown in third (Starting date) and fourth (End date) columns. There are a few false positive events, the ratio for the critical event August 2011 fall (Serial No 2) and IPO Facebook Debut (Serial No 8) show the value higher than 0.4.}
\label{table:ellipsoid_jpn}
\end{table}

\section{Conclusions and future outlook}
We have presented a brief overview of the evolution of the states of financial markets based on the similarity of correlation matrices obtained over $T=40$ and $20$ days epochs with shifts ranging from the entire length of the epoch using daily shifts.  The adjusted closing data of the S\&P 500 and the Nikkei 225 indices are appropriately purified.
The line of ideas starts with the work by Munnix et. al. \cite{munnix2012identifying} in 2012, where they proposed to group similar correlation structures as ``market state” and showed the evolution of the financial market through these market states. They used a top-bottom $k$-means clustering technique, which was later replaced by minimization of intra-cluster size while requiring stability of noise reduction in order to determine the optimal number of clusters. We then describe a more recent development to take advantage of some remaining freedom of choice by searching for desirable properties of the transition matrix between states.
It is important at this stage, to mention that noise reduction is carried out at the level of the correlation matrix rather than at the level of individual time series. We briefly discuss the power map introduced and used extensively by the group of Guhr et. al. \cite{guhr2003,schafer2010power}. The amusing interpretation of this method as an artificial lengthening of the time series \cite{guhr2003} is revisited by analyzing its effects in synthetic data obtained from correlated Wishart ensembles \cite{pharasi2020}.

The line of argument leads up to recent work detailing the use for risk assessment \cite{pharasi2020dynamics}. It brings a better perspective to hedge against the market’s critical events without a prohibitive cost. By using the same overlapping epoch of length 20 days but the shifts of $1$ day, the results showed significantly improved statistics as compared to 10 days shifts. Using this criterion over the period of 2006-2019, the S\&P 500 and Nikkei 225 markets is characterized into five market states with $\epsilon=0.7$ and seven market states with $\epsilon=0$ (no suppression).

Further, we studied the cluster analysis of market states constructed from averaged intra- and inter-sectorial matrices to identify the market dynamics. Using the same data over the period 2016-2019, the results for S\&P 500 and Nikkei 225 are statistically similar to the one obtained from the stock analysis. We showed that the better choice of transition matrix over a minimum of ${\sigma_{d_{intra}}}$. In the Japanese case, the preference is given to a transition matrix that avoids large jumps from lower states to the critical state. Using this approach, we have a better transition matrix and similar distribution of clusters as obtained from the analysis of the stocks. Note that in this approach, we do not perform the clustering of a correlation matrix. A simple-minded approach averaging the series corresponding to each sector yields a different result that does not seem satisfactory.
Finally, we try to look a little into the future, by inspecting the time dynamics in the dimensionally scaled space of the correlation matrices in a representation dimensionally scaled to three dimensions, which seems quite interesting, though no conclusions emerge. This leads to an alternative proposition using the market states as the domains on which  symbolic dynamics can be built and this also seems like a promising lane to expand the research. A more detailed analysis of the structure of each cluster is an another interesting topic of future research.

\section*{Acknowledgment}
The authors are grateful to   Francois Leyvraz for their critical inputs and suggestions. H.K.P., P.M. and S.S. are grateful for financial support provided by UNAM-DGAPA and CONACYT Proyecto Fronteras 952. T.H.S. and H.K.P. acknowledge the support grant by CONACYT Proyecto Fronteras 201, UNAM-DGAPA-PAPIIT AG100819 and IN113620. T.H.S. and H.K.P. also acknowledge computing support under project LANCAD-UNAM-DGTIC-016. 
\bibliography{bib_all_corrected}%

\begin{thebibliography}{10}
\expandafter\ifx\csname url\endcsname\relax
  \def\url#1{\texttt{#1}}\fi
\expandafter\ifx\csname urlprefix\endcsname\relax\def\urlprefix{URL }\fi
\expandafter\ifx\csname doiprefix\endcsname\relax\def\doiprefix{DOI }\fi
\providecommand{\bibinfo}[2]{#2}
\providecommand{\eprint}[2][]{\url{#2}}

\bibitem{munnix2012identifying}
\bibinfo{author}{M{\"u}nnix, M.~C.} \emph{et~al.}
\newblock \bibinfo{title}{Identifying states of a financial market}.
\newblock \emph{\bibinfo{journal}{Scientific Reports}}
  \textbf{\bibinfo{volume}{2}}, \bibinfo{pages}{644} (\bibinfo{year}{2012}).
\newblock \doiprefix 10.1038/srep00644.

\bibitem{rinn2015dynamics}
\bibinfo{author}{Rinn, P.}, \bibinfo{author}{Stepanov, Y.},
  \bibinfo{author}{Peinke, J.}, \bibinfo{author}{Guhr, T.} \&
  \bibinfo{author}{Sch{\"a}fer, R.}
\newblock \bibinfo{title}{Dynamics of quasi-stationary systems: Finance as an
  example}.
\newblock \emph{\bibinfo{journal}{EPL (Europhysics Letters)}}
  \textbf{\bibinfo{volume}{110}}, \bibinfo{pages}{68003}
  (\bibinfo{year}{2015}).
\newblock \doiprefix 10.1209/0295-5075/110/68003.

\bibitem{pharasi2020dynamics}
\bibinfo{author}{Pharasi, H.~K.}, \bibinfo{author}{Seligman, E.},
  \bibinfo{author}{Sadhukhan, S.} \& \bibinfo{author}{Seligman, T.~H.}
\newblock \bibinfo{title}{Dynamics of market states and risk assessment}.
\newblock \emph{\bibinfo{journal}{arXiv preprint arXiv:2011.05984}}
  (\bibinfo{year}{2020}).

\bibitem{Pharasi_2018}
\bibinfo{author}{Pharasi, H.~K.} \emph{et~al.}
\newblock \bibinfo{title}{Identifying long-term precursors of financial market
  crashes using correlation patterns}.
\newblock \emph{\bibinfo{journal}{New Journal of Physics}}
  \textbf{\bibinfo{volume}{20}}, \bibinfo{pages}{103041}
  (\bibinfo{year}{2018}).
\newblock \doiprefix 10.1088/1367-2630/aae7e0.

\bibitem{Pharasi_2019}
\bibinfo{author}{Pharasi, H.~K.}, \bibinfo{author}{Sharma, K.},
  \bibinfo{author}{Chakraborti, A.} \& \bibinfo{author}{Seligman, T.~H.}
\newblock \bibinfo{title}{Complex market dynamics in the light of random matrix
  theory}.
\newblock In \emph{\bibinfo{booktitle}{New Perspectives and Challenges in
  Econophysics and Sociophysics}}, \bibinfo{pages}{13--34}
  (\bibinfo{publisher}{Springer International Publishing},
  \bibinfo{address}{Cham}, \bibinfo{year}{2019}).

\bibitem{chetalova2015zooming}
\bibinfo{author}{Chetalova, D.}, \bibinfo{author}{Sch{\"a}fer, R.} \&
  \bibinfo{author}{Guhr, T.}
\newblock \bibinfo{title}{Zooming into market states}.
\newblock \emph{\bibinfo{journal}{Journal of Statistical Mechanics: Theory and
  Experiment}} \textbf{\bibinfo{volume}{2015}}, \bibinfo{pages}{P01029}
  (\bibinfo{year}{2015}).

\bibitem{stepanov2015stability}
\bibinfo{author}{Stepanov, Y.}, \bibinfo{author}{Rinn, P.},
  \bibinfo{author}{Guhr, T.}, \bibinfo{author}{Peinke, J.} \&
  \bibinfo{author}{Sch{\"a}fer, R.}
\newblock \bibinfo{title}{Stability and hierarchy of quasi-stationary states:
  financial markets as an example}.
\newblock \emph{\bibinfo{journal}{Journal of Statistical Mechanics: Theory and
  Experiment}} \textbf{\bibinfo{volume}{2015}}, \bibinfo{pages}{P08011}
  (\bibinfo{year}{2015}).
\newblock \doiprefix 10.1088/1742-5468/2015/08/P08011.

\bibitem{guhr2020exact}
\bibinfo{author}{Guhr, T.} \& \bibinfo{author}{Schell, A.}
\newblock \bibinfo{title}{Exact multivariate amplitude distributions for
  non-stationary gaussian or algebraic fluctuations of covariances or
  correlations} (\bibinfo{year}{2020}).
\newblock \eprint{2011.07570}.

\bibitem{Heckens_2020}
\bibinfo{author}{Heckens, A.~J.}, \bibinfo{author}{Krause, S.~M.} \&
  \bibinfo{author}{Guhr, T.}
\newblock \bibinfo{title}{Uncovering the dynamics of correlation structures
  relative to the collective market motion}.
\newblock \emph{\bibinfo{journal}{Journal of Statistical Mechanics: Theory and
  Experiment}} \textbf{\bibinfo{volume}{2020}}, \bibinfo{pages}{103402}
  (\bibinfo{year}{2020}).

\bibitem{Meudt_2015}
\bibinfo{author}{Meudt, F.}, \bibinfo{author}{Theissen, M.},
  \bibinfo{author}{Schäfer, R.} \& \bibinfo{author}{Guhr, T.}
\newblock \bibinfo{title}{Constructing analytically tractable ensembles of
  stochastic covariances with an application to financial data}.
\newblock \emph{\bibinfo{journal}{Journal of Statistical Mechanics: Theory and
  Experiment}} \textbf{\bibinfo{volume}{2015}}, \bibinfo{pages}{P11025}
  (\bibinfo{year}{2015}).

\bibitem{guhr2015non}
\bibinfo{author}{Guhr, T.}
\newblock \bibinfo{title}{Non-stationarity in financial markets: Dynamics of
  market states versus generic features.}
\newblock \emph{\bibinfo{journal}{Acta Physica Polonica B}}
  \textbf{\bibinfo{volume}{46}} (\bibinfo{year}{2015}).

\bibitem{husain2020identifying}
\bibinfo{author}{Husain, S.~S.}, \bibinfo{author}{Sharma, K.},
  \bibinfo{author}{Kukreti, V.} \& \bibinfo{author}{Chakraborti, A.}
\newblock \bibinfo{title}{Identifying the global terror hubs and vulnerable
  motifs using complex network dynamics}.
\newblock \emph{\bibinfo{journal}{Physica A: Statistical Mechanics and Its
  Applications}} \textbf{\bibinfo{volume}{540}}, \bibinfo{pages}{123113}
  (\bibinfo{year}{2020}).

\bibitem{wang2020quasi}
\bibinfo{author}{Wang, S.}, \bibinfo{author}{Gartzke, S.},
  \bibinfo{author}{Schreckenberg, M.} \& \bibinfo{author}{Guhr, T.}
\newblock \bibinfo{title}{Quasi-stationary states in temporal correlations for
  traffic systems: Cologne orbital motorway as an example}.
\newblock \emph{\bibinfo{journal}{Journal of Statistical Mechanics: Theory and
  Experiment}} \textbf{\bibinfo{volume}{2020}}, \bibinfo{pages}{103404}
  (\bibinfo{year}{2020}).

\bibitem{scheffer_2009}
\bibinfo{author}{Scheffer, M.}
\newblock \emph{\bibinfo{title}{Critical Transitions in Nature and Society}}.
\newblock Princeton Studies in Complexity (\bibinfo{publisher}{Princeton
  University Press}, \bibinfo{year}{2009}).

\bibitem{scheffer_2012}
\bibinfo{author}{Scheffer, M.} \emph{et~al.}
\newblock \bibinfo{title}{Anticipating critical transitions}.
\newblock \emph{\bibinfo{journal}{Science}} \textbf{\bibinfo{volume}{338}},
  \bibinfo{pages}{344--348} (\bibinfo{year}{2012}).
\newblock \doiprefix 10.1126/science.1225244.

\bibitem{May_2008}
\bibinfo{author}{May, R.~M.}, \bibinfo{author}{Levin, S.~A.} \&
  \bibinfo{author}{Sugihara, G.}
\newblock \bibinfo{title}{Ecology for bankers}.
\newblock \emph{\bibinfo{journal}{Nature}} \textbf{\bibinfo{volume}{451}},
  \bibinfo{pages}{893–--895} (\bibinfo{year}{2008}).
\newblock \doiprefix 10.1038/451893a.

\bibitem{Sornette_2004}
\bibinfo{author}{Sornette, D.}
\newblock \emph{\bibinfo{title}{Why Stock Markets Crash: Critical Events in
  Complex Financial Systems}} (\bibinfo{publisher}{Princeton University Press},
  \bibinfo{year}{2004}).

\bibitem{weiss2004social}
\bibinfo{author}{Weiss, R.~A.} \& \bibinfo{author}{McMichael, A.~J.}
\newblock \bibinfo{title}{Social and environmental risk factors in the
  emergence of infectious diseases}.
\newblock \emph{\bibinfo{journal}{Nature Medicine}}
  \textbf{\bibinfo{volume}{10}}, \bibinfo{pages}{S70--S76}
  (\bibinfo{year}{2004}).

\bibitem{wang2017unification}
\bibinfo{author}{Wang, W.}, \bibinfo{author}{Tang, M.},
  \bibinfo{author}{Stanley, H.~E.} \& \bibinfo{author}{Braunstein, L.~A.}
\newblock \bibinfo{title}{Unification of theoretical approaches for epidemic
  spreading on complex networks}.
\newblock \emph{\bibinfo{journal}{Reports on Progress in Physics}}
  \textbf{\bibinfo{volume}{80}}, \bibinfo{pages}{036603}
  (\bibinfo{year}{2017}).

\bibitem{kawamura2012stat}
\bibinfo{author}{Kawamura, H.}, \bibinfo{author}{Hatano, T.},
  \bibinfo{author}{Kato, N.}, \bibinfo{author}{Biswas, S.} \&
  \bibinfo{author}{Chakrabarti, B.~K.}
\newblock \bibinfo{title}{Statistical physics of fracture, friction, and
  earthquakes}.
\newblock \emph{\bibinfo{journal}{Rev. Mod. Phys.}}
  \textbf{\bibinfo{volume}{84}}, \bibinfo{pages}{839--884}
  (\bibinfo{year}{2012}).
\newblock \urlprefix\url{https://link.aps.org/doi/10.1103/RevModPhys.84.839}.
\newblock \doiprefix 10.1103/RevModPhys.84.839.

\bibitem{muller2011evolution}
\bibinfo{author}{M{\"u}ller, M.~F.} \emph{et~al.}
\newblock \bibinfo{title}{Evolution of genuine cross-correlation strength of
  focal onset seizures}.
\newblock \emph{\bibinfo{journal}{Journal of Clinical Neurophysiology}}
  \textbf{\bibinfo{volume}{28}}, \bibinfo{pages}{450--462}
  (\bibinfo{year}{2011}).

\bibitem{kwapien2012physical}
\bibinfo{author}{Kwapie{\'n}, J.} \& \bibinfo{author}{Dro{\.z}d{\.z}, S.}
\newblock \bibinfo{title}{Physical approach to complex systems}.
\newblock \emph{\bibinfo{journal}{Physics Reports}}
  \textbf{\bibinfo{volume}{515}}, \bibinfo{pages}{115--226}
  (\bibinfo{year}{2012}).

\bibitem{torgerson1952multidimensioal}
\bibinfo{author}{Torgerson, W.}
\newblock \bibinfo{title}{Multidimensional scaling: I. theory and method}.
\newblock \emph{\bibinfo{journal}{Psychometrika}}
  \textbf{\bibinfo{volume}{17}}, \bibinfo{pages}{401--419}
  (\bibinfo{year}{1952}).
\newblock \doiprefix 10.1007/BF02288916.

\bibitem{kerner2009introduction}
\bibinfo{author}{Kerner, B.~S.}
\newblock \emph{\bibinfo{title}{Introduction to modern traffic flow theory and
  control: the long road to three-phase traffic theory}}
  (\bibinfo{publisher}{Springer Science \& Business Media},
  \bibinfo{year}{2009}).

\bibitem{Yahoo_finance}
\bibinfo{title}{Yahoo finance database}.
\newblock \bibinfo{howpublished}{\url{https://finance.yahoo.com/}}
  (\bibinfo{year}{2020}).
\newblock \bibinfo{note}{Accessed on 7th January, 2020 for S\&P 500 and 14th
  January, 2020 for Nikkei 225.}

\bibitem{plerou2000random}
\bibinfo{author}{Plerou, V.}, \bibinfo{author}{Gopikrishnan, P.},
  \bibinfo{author}{Rosenow, B.}, \bibinfo{author}{Amaral, L.~N.} \&
  \bibinfo{author}{Stanley, H.~E.}
\newblock \bibinfo{title}{A random matrix theory approach to financial
  cross-correlations}.
\newblock \emph{\bibinfo{journal}{Physica A: Statistical Mechanics and its
  Applications}} \textbf{\bibinfo{volume}{287}}, \bibinfo{pages}{374--382}
  (\bibinfo{year}{2000}).
\newblock \doiprefix 10.1016/S0378-4371(00)00376-9.

\bibitem{pandey2010correlated}
\bibinfo{author}{Pandey, A.} \emph{et~al.}
\newblock \bibinfo{title}{Correlated wishart ensembles and chaotic time
  series}.
\newblock \emph{\bibinfo{journal}{Physical Review E}}
  \textbf{\bibinfo{volume}{81}}, \bibinfo{pages}{036202}
  (\bibinfo{year}{2010}).

\bibitem{Guhr_2003}
\bibinfo{author}{Guhr, T.} \& \bibinfo{author}{K{\"a}lber, B.}
\newblock \bibinfo{title}{A new method to estimate the noise in financial
  correlation matrices}.
\newblock \emph{\bibinfo{journal}{Journal of Physics A: Mathematical and
  General}} \textbf{\bibinfo{volume}{36}}, \bibinfo{pages}{3009}
  (\bibinfo{year}{2003}).
\newblock \doiprefix 10.1088/0305-4470/36/12/310.

\bibitem{vinayak_2014}
\bibinfo{author}{Vinayak} \& \bibinfo{author}{Seligman, T.~H.}
\newblock \bibinfo{title}{Time series, correlation matrices and random matrix
  models}.
\newblock In \emph{\bibinfo{booktitle}{AIP Conference Proceedings}}, vol.
  \bibinfo{volume}{1575}, \bibinfo{pages}{196--217}
  (\bibinfo{organization}{AIP}, \bibinfo{year}{2014}).

\bibitem{Chakraborti_2020}
\bibinfo{author}{Chakraborti, A.} \emph{et~al.}
\newblock \bibinfo{title}{Emerging spectra characterization of catastrophic
  instabilities in complex systems}.
\newblock \emph{\bibinfo{journal}{New Journal of Physics}}
  \textbf{\bibinfo{volume}{22}}, \bibinfo{pages}{063043}
  (\bibinfo{year}{2020}).

\bibitem{vinayakpre2013}
\bibinfo{author}{Vinayak}, \bibinfo{author}{Sch\"afer, R.} \&
  \bibinfo{author}{Seligman, T.~H.}
\newblock \bibinfo{title}{Emerging spectra of singular correlation matrices
  under small power-map deformations}.
\newblock \emph{\bibinfo{journal}{Physical Review E}}
  \textbf{\bibinfo{volume}{88}}, \bibinfo{pages}{032115}
  (\bibinfo{year}{2013}).
\newblock \doiprefix 10.1103/PhysRevE.88.032115.

\bibitem{schafer2010power}
\bibinfo{author}{Sch{\"a}fer, R.}, \bibinfo{author}{Nilsson, N.~F.} \&
  \bibinfo{author}{Guhr, T.}
\newblock \bibinfo{title}{Power mapping with dynamical adjustment for improved
  portfolio optimization}.
\newblock \emph{\bibinfo{journal}{Quantitative Finance}}
  \textbf{\bibinfo{volume}{10}}, \bibinfo{pages}{107--119}
  (\bibinfo{year}{2010}).

\bibitem{laloux1999noise}
\bibinfo{author}{Laloux, L.}, \bibinfo{author}{Cizeau, P.},
  \bibinfo{author}{Bouchaud, J.-P.} \& \bibinfo{author}{Potters, M.}
\newblock \bibinfo{title}{Noise dressing of financial correlation matrices}.
\newblock \emph{\bibinfo{journal}{Physical Review Letters}}
  \textbf{\bibinfo{volume}{83}}, \bibinfo{pages}{1467--1470}
  (\bibinfo{year}{1999}).
\newblock \doiprefix 10.1103/PhysRevLett.83.1467.

\bibitem{guhr2003}
\bibinfo{author}{Guhr, T.} \& \bibinfo{author}{K{\"a}lber, B.}
\newblock \bibinfo{title}{A new method to estimate the noise in financial
  correlation matrices}.
\newblock \emph{\bibinfo{journal}{Journal of Physics A: Mathematical and
  General}} \textbf{\bibinfo{volume}{36}}, \bibinfo{pages}{3009}
  (\bibinfo{year}{2003}).
\newblock \doiprefix 10.1088/0305-4470/36/12/310.

\bibitem{pharasi2020}
\bibinfo{author}{Pharasi, H.~K.}, \bibinfo{author}{Seligman, E.} \&
  \bibinfo{author}{Seligman, T.~H.}
\newblock \bibinfo{title}{Market states: A new understanding}.
\newblock \emph{\bibinfo{journal}{arXiv preprint arXiv:2003.07058}}
  (\bibinfo{year}{2020}).

\end{thebibliography}

\end{document}